\documentclass[prl,twocolumn,showpacs]{revtex4}%
\usepackage{amsfonts}
\usepackage{amsmath}
\usepackage{amssymb}
\usepackage{graphicx}%
\providecommand{\U}[1]{\protect\rule{.1in}{.1in}}

\begin{document}
\title{\textbf{Towards a unified theory of the fundamental physical interactions
based on the underlying geometric structure of the tangent bundle}}

\author{Joachim Herrmann}
\affiliation{Max Born Institute, Max Born Str. 2a , D12489 Berlin, Germany}
\email{jherrman@mbi-berlin.de}

\begin{abstract}
This paper pursues the hypothesis that the tangent bundle (TB) with the
central extended little groups of the SO(3,1) group as gauge group is the
underlying geometric structure for a unified theory of the fundamental
physical interactions. Based on this hypothesis as a first step recently I
presented a generalized theory of electroweak interaction \ which includes
hypothetical dark matter particles (Eur. Phys. J C 79, 779 (2019). The
vertical Laplacian of the tangent bundle possesses the same form as the
Hamiltonian of a 2D semiconductor quantum Hall system. This explains
fractional charge quantization of quarks and the existence of lepton and quark
families. As will be shown the SU(3) colour symmetry for strong interaction
arises in the TB as an emergent symmetry similar as Chern-Simon gauge
symmetries in quantum Hall systems. This predicts a signature of quark
confinement as an universal large-scale property of the Chern-Simon fields and
induces a new understanding of the vacuum as the ground state occupied with a
condensate of quark-antiquark pairs. The gap for quark-antiquark pairing is
calculated in the mean-field approximation which allows a numerical estimation
of the characteristic parameters of the vacuum such as its chemical potential,
the quark condensation parameter and the vacuum energy. Note that previously a
gauge theoretical understanding of gravity has been achieved by considering
the translation group T(3,1) in the TB as gauge group. Therefore the theory
presented here can be considered as a new type of unified theory for all known
fundamental interactions linked with the geometrization program of physics.

\end{abstract}
\maketitle

\bigskip\textbf{1. Introduction}

The developement of the Standard Model (SM) of elementary particles was a
great triumph of modern physics. However despite its enormous success in
explaining a large number of experiments, it cannot be considered as the
ultimate theory of particle physics. The SM is incomplete and contains several
open problems that cannot be solved on the basis of this model, as well as
phenomenological elements that have not yet been explained on a microscopic
basis. A fundamental problem in the SM is the question of the physical origin
of the internal symmetry groups assumed a priory for phenomenological reasons.
Several other unanswered questions are the mysterious existence of three
families of leptons and quarks (also called generations) that differ only by
their masses and the origin of the hierarchy of fermion masses, the
microscopic understanding of the spontaneous symmetry breaking by the Higgs
mechanism and the hierarchy problem, the lack of understanding of fractional
quark charges, the missing explanation of dark matter and dark energy, and others.

There have been many attempts to formulate a unified theory beyond the SM that
could solve the above puzzles. In Grand Unified theories, the electroweak and
strong interactions are embedded in a larger gauge group, such as the SU(5)
group first proposed by Georgi and Glashow \cite{G1} or in the SO(10) group
\cite{G2}. There have also been some attempts to understand the origin of
families in the SM by using a so-called family symmetry \cite{G3,G4}. Despite
many subsequent attempts, no unified model exists that is able to solve all
these problems on a unified basis and none of them are considered to be
universally accepted.

The SM of particle physics is based on the gauge principle, with local
symmetry transformations of the variables of the phenomenologically determined
continuos gauge groups $SU(2)\otimes U(1$) in electroweak interaction and
$SU(3$) in strong interaction (quantum chromodynamics) acting in an "internal
space" related to the symmetry group. The geometric or physical origin of the
internal symmetries given by the gauge transformations remains unknown. Here
we follow the hypothesis that the gauge principle is not a complete
independent postulate but\ is connected with a deeper principle which allows
one to determine the internal symmetries without phenomenological assumptions.
A possible way to realize this aim is by revisiting the geometrization program
of unified field theory in the 1920s to find a more general geometric
structure than the Rieman geometry of the world spacetime manifold for the
unification of gravity with electromagnetism and its combination with the
knowledge in modern particle physics. This requires a synthesis of the
principle of general relativity and gauge transformations in the frame of an
united geometric structure. Note that the general type of such geometric
structure has been recognized since the 1960s by the discovery of a formal
equivalence of gauge theories with the mathematical formalism of fiber bundles
\cite{r1,r2,r3,r4,r5,r6,r7}. In the fiber bundle interpretation of gauge
theories, the gauge potentials are understood as a geometric entity, the
connection on the principal bundle; matter fields are described by the
associated fiber bundles and gauge transformations are identified as
transformations of the fiber variables along the fiber axis at fixed space
points \textbf{x}. But up to now one key question remained unanswered: which
specific fibre manifold could be the basis for a unified theory of the
fundamental forces? In previous studies the transformation groups of the
fibers were taken from the phenomenologically determined gauge groups of the
SM. Therefore the fiber bundle approach mainly delivered a geometrization and
reinterpretation of the gauge potentials but could not be used as a bridge to
a theory beyond the SM.

The present paper pursues the hypothesis \cite{I1} that the fundamental
interactions\textit{\ }are linked with\textit{\ } geometric symmetries in the
underlying extended geometric\textbf{\ } structure of the most fundamental
fiber bundle- the tangent bundle. A tangent bundle is a more general manifold
than the pseudo-Rieman manifold $M$ in the GTR. It associates to every point
$x$ of the space-time manifold $M$ a 4-dimensional tangent space $T_{x}(M)$
which is the set of all tangent vectors at point $x. $ In $T_{x}(M)$ one can
define the addition of tangent vectors and multiplication between a tangent
vector and a real number. The union of all tangent spaces at all points $x$ of
the space-time manifold $M$ is called the tangent bundle $TM$. Tangent vectors
in $T_{x}(M)$ can be transformed by the special affine group $G=SO(3,1)\rtimes
T(3,1)$, but there is generally no natural choice of isomorphism between
$T_{x}(M)$ and $T_{y}(M)$ for $x\neq y.$ Any point can be mapped to the base
manifold by a projection map $\pi.$ In a Riemann geometry the differential
calculus requires the existence of a connection between the tangent spaces of
neighboring points. Such isomorphism can be constructed by the parallel
transport using the Levi-Civita connection which depends on the metric of the
space-time manifold. In the more general geometric structure of fiber bundles
the connection is independent on the metric and can be axiomatically defined
as a matrix-valued 1-form \cite{TG1,TG2}.

The geometry of the TB is closely linked to the conceptional basis of gravity
theories and its extensions to gravity gauge theories. Considerable efforts
has been made in the construction of a gauge theory for gravity by analogy
with gauge theories in the SM of particle physics. The theory proposed by
Kibble (\cite{P1} and Sciama (\cite{p2}) marks the earliest attempts to
formulate a gauge theory of gravity based on the localization of the Poincare
group as gauge group (for reviews see \cite{P2,P3,P4}). The Kibble-Sciama
theory incorporates besides curvature, Cartans torsion as a dynamical variable
and predicts that the torsion tensor is related to the spin density through a
linear, algebraic equation, so that torsion does not propagate. This theory
leads to a hypothetical generalized gravity theory, the Einstein-Cartan
gravity theory. Although the present paper consider the same mathematical
group for a gauge-theoretical construction, the theory presented differs
significantly from the Kibble-Sciama theory in the main technical details, in
the aim and in the decribed physical object. The Kibble- Sciama theory uses a
heuristic scheme within the Minkowski space of special relativity with the
Poincare group as a global symmetry for coordinate transformations in the flat
Minkowsky spacetime manifold. Applying the gauge principle to this symmetry
leads to Rieman-Cartan geometry with curvature and torsion due to the
intrinsic spin of matter. It is not apparent from this procedure what is
actually transformed by the localized Poincare transformation. Particularly
peobematic is the interpretation of local translations which act not only on
fields, but also on spacetime points. A precise analysis of the gauge content
reveals certain structural differences with repect to the other SM gauge
transformations. As many authors have remarked (see e.g.(\cite{r2,t2,g5}) more
advantagous is to work in the framework of fiber bundles, which avoids various
pitfalls and confusions arising from mixing up of basic transformations in the
tangent space and that of general coordinate transformations in the spacetime
manifold. So far the theory of Kibble and Sciama has not found general
acceptance as a viable theory of gravity because there has been no
experimental evidence supporting the excistence of gravitational torsion.

There exists a second class of a gauge theory of gravity which has found large
attention by considering instead the full Poincare group the translation
sub-group $T(3,1)$ in the tangent bundle as gauge group
\cite{t1,t2,g5,t3,t4,t5}\ leading to the teleparallel gravity theory. In this
theory the connection takes the form of the Weizenb\"{o}ck connection, the
curvature is identical zero and gravity is decribed in terms of torsion
instead of the curvature. The action only differ by a boundary term from the
Einstein-Hilbert action and thus teleparallel gravity and Einsteins theory of
gravity are classical equivalently. Consequently this theory is just as
experimentally viable as Einsteins gravity theory.

There is a further essential point in th\'{\i}s context in that the Poincare
group is not semi-simple. The action of this group on a vector space is not
transitive, but the vector space decomposes into different orbits in which the
group acts transitively. Wigner developed the method of induced
representations to find the unitary representation of the Poincare group by
using projective representations. For the case of simply\ connected groups
like the rotation group $SO(3$) projective representations are obtained by
replacing the group $SO(3)$ by its universal cover $SU(2)$. However the
Euclidean group $E(2)$ is not semi-simple and the covering group
$\widetilde{E}(2)$ is not sufficient. One has to use a larger group: the
universal central extension $E^{c}(2).$

In the present paper it is assumed that gravity can be described by the
translational gauge theory with the structure group $T(3,1)$ based on the
translational connection as gauge field. However, the restriction of the
symmetry group $G=SO(3,1)\rtimes T(3,1)$ in the TB to the subgroup $T(3,1)$
seems to be somewhat dissatisfying and introduced several controversies if
spinor fields are present (see e.g. \cite{D1,D2,D3}). One can avoid these
problems if the spin-connection is not only related with inertia. This raises
the question of the physical meaning of the other subgroup $SO(3,1)$ which is
not connected with gravity. The basic hypothesis in the present manuscript is
that sub-group $SO(3,1)$ is related with the other fundamental interactions.
As a first step based on this hypothesis in \cite{I1} a generalized theory of
electroweak interaction without a priory assumption of a phenomenological
determined gauge group but with internal (gauge) symmetries arising from
geometric transformations of tangent vectors along the tangent vector axis is
presented. The little groups $SU(2)\otimes E^{c}(2)$ of the non-transitive
group $SO(1,3)$ were defined as structure groups (gauge groups) of the fiber
bundle. The eigenfunctions of the Laplacian on this product group yields the
known internal quantum numbers of iso-spin and hypercharge, but in addition
the $E^{c}$-charge $\varkappa$ and the family quantum number $n$. The
connection coefficients (gauge potentials) describe the known SM gauge bosons
but additional extra gauge bosons are predicted (which can be interpreted as
dark vector bosons). Besides the SM particles candidate stable and instable
dark matter fermions and dark matter scalars without additional a priory
phenomenological model assumptions arise from a common origin with SM
particles. On the other hand the color group $SU(3)$ of QCD cannot be
described as a geometric symmetry in an analog way as the $SU(2)\otimes
E^{c}(2)$ group. However the TB delivers an interesting link which could pave
the path for the inclusion of the color symmetry $SU(3)$ into a unified theory
based on the TB geometry. A surprising feature in strong interaction is the
fact that quarks carry fractional electric charges similar to electrons in a
quantum Hall system. Up to now this analogy in the property of quarks and
electrons in a quantum Hall system has not found a convincing explanation. On
the other hand, the eigenfunctions of the Laplacian of the $E^{c}(2)$ group
have the same form as the solution of the 2D Schr\"{o}dinger equation for
electrons in a perpendicular external magnetic field \cite{H1,H4,H5}.
Combining all tangent fibers at all spacetime points, the vertical (internal)
bundle Laplacian of the TB obtains a form analogous as the multi-particle
Hamiltonian of a 2D quantum Hall system. The eigenstates in a quantum Hall
system are called Landau levels, they show the same form as the solutions with
internal quantum numbers (IQN) $n$ of the Laplacian of the $E^{c}(2)$ group
(here denoted with TB Landau levels,TB-LL). In the SM members of different
families have identical IQNs and properties except of its masses. In contrast
in the TB the existence of three families of leptons and quarks can be
distinguished by the different family numbers $n=1,2,3$ while the vacuum state
(ground state) carry the TB family quantum number $n=0$. The analogy of
fractional quark charges with the quantum Hall Effect (QHE) requires the
additional hypothesis that the vacuum with the IQN $n=0$ is filled with
seeleptons and composite seequarks and all higher levels $n=1,2,3$ are empty.
Furthers an energy gap exists between the lowest TB-LL and a higher TB-LL. If
a seequark is excited into a higher TB-LL it creates an excited state which
will be denoted as valence quark and leaves a hole in the old state. A lepton
or quark hole carry the opposite hypercharge and opposite isospin IQN, but a
positive energy and can be interpreted as anti-particle. Taking into account
the three isospin components of quarks with $I_{3}=1/2,0,-1/2$ the vertical
Laplacian obtains the analog form as the Hamiltonian of a three-layer quantum
Hall system. In this approach, effective gauge fields (denoted as Chern-Simon
fields) with a local SU(3) symmetry arise in the vertical TB Laplacian in an
internal way. An emergent phenomenon is a collective effect of a large number
of particles that cannot be deduced from the microscopic theory in a rigorous
way \cite{H6}. The fractional QHE\ is a prototype of such phenomenon.

In the model of a completely filled vacuum state due to the presence of
attractive interaction by gluon exchange the vacuum is instable with respect
to the formation of a quark condensate caused by the pairing of quarks with
anti-quarks. This phenomenon is analogous to exciton condensation in solid
states \cite{E1,E2,E3,E4} where pairs of electrons and holes form a condensate
due to the weak attractive force. The condensed TB vacuum is characterized as
a completely filled state with a finite particle density, therefore we have to
introduce the Fermi-energy of the vacuum $E_{Fermi}$ or a vacuum chemical
potential $\mu_{vac}$ with $\mu_{vac}\simeq E_{Fermi}$ . To explore the
dynamics of quark condensation by quark-antiquark pairing we will calculate
the energy gap in the mean-field approximation of the relativistic Hamiltonian
formalism taking into account the vacuum expectation for quark-antiquark
pairing. The numerical parameter of the energy gap $\Delta_{gap}$ is
determined by a relation of the quark condensation parameter with the gap
parameter and the chemical potential $\mu_{vac}$.

The paper is organized as follows. In chapter 2 fundamentals of differential
geometry on the TB are briefly described. Chapter 3 describes how the color
$SU(3)$ symmetry for strong interaction arises as an emergent symmetry similar
as Chern-Simon gauge fields in quantum Hall systems. In chapter 4 the
condensed vacuum structure in the TB with a quark condensate is described and
the gap equation for quark-antiquark pairing is derived.

\textbf{2. Basics of the tangent bundle \ geometry and central extensions of
the little groups}

The tangent bundle is one of the most important concept in differential
geometry on curved manifolds. A tangent vector can be defined as \ linear
differential operators $v_{P}f(x(t)=\frac{d}{dt}f(x(t)_{P}$ which acts on
functions on a spacetime manifold $M$ tangent to the curve x(t) at the space
point $P$. The set of all tangent vectors $v_{P}$ spanned by frame vectors in
a given basis is the tangent space $T_{P}(M).$ This means tangent vectors at
the point $P$ are placed in their own space $T_{P}(M)$. The tangent bundle TM
is the disjoint union of all tangent spaces at all points $P$:
\begin{align}
TM  &  :=\bigcup{}_{P\in M}T_{P}(M)\tag{2.1}\\
&  =\{(x,v):x\in M,v\in T_{x}(M)\},\nonumber
\end{align}
In coordinate description a point in the tangent bundle is given by the pairs
$X=(x,v)$ with $x=(x_{0,}x_{1,}x_{2,}x_{3})$ as the coordinate of the
spacetime manifold and $v=(v_{0,}v_{1,}v_{2,}v_{3})$ the coordinate of tangent
vectors, i.e. the TB is a 8-dimensional manifold. We introduce tetrads
$e_{a}^{\mu}(x)$ which form an orthonormal basis $g_{\mu\nu}(x)e_{a}^{\mu
}(x)e_{b}^{\nu}(x)=\eta_{ab}$ where $g_{\mu\nu}(x)$ is the metric of the
spacetime manifold and $\eta_{ab}=diag(-1,1,1,1)$ the metric of the Minkowski
\ space. The four tetrads $e_{a}=e_{a}^{\mu}(x)\partial_{\mu}$ form the basis
for the tangent space at each spacetime point $x$. The subscript $a.b$ numbers
the vectors ($a,b=0,1,2,3$) and $\mu=0,1,2,3$ their components in the
coordinate basis $e_{\mu}=$ $\partial_{\mu}$ ($\mu=0,1,2,3).$ Each vector $v$
described in the coordinate basis $e_{\mu}=$ $\partial_{\mu}$ can be expressed
by a vector with respect to the frame basis $e_{a}$ according to the rule
$v^{\nu}=e_{a}^{\nu}(x)v^{a}$. Cotangent 1-forms $\ \theta^{a}=\theta_{\mu
}^{a}dx^{\mu}$ can be defined satisfying the orthogonality relation
$e_{a}^{\mu}(x)\theta_{\nu}^{a}(x)=\delta_{\nu}^{\mu}$. The tetrads induce a
spacetime pseudo-Rieman metric $g_{\mu\nu}(x)=\theta_{\mu}^{a}\theta_{\nu}%
^{b}$ $\eta_{ab}.$ The scalar product of a vector and a co-vector is defined
as%
\begin{equation}
(v,u)=g_{\mu\nu}(x)v^{\mu}u^{\nu}=\eta_{ab}v^{a}u^{b}, \tag{2.2}%
\end{equation}

The scalar product (2.2) is the governing structure relation of the tangent
bundle, its invariance with respect to certain transformations determines the
geometry of the TB. There exist two types of transformations which leaves
(2.2) invariant. First, general spacetime coordinate transformations $x^{\mu
}\rightarrow y^{\mu}=y^{\mu}(x)$ with vectors which are transformed as
$v^{\prime\mu}(x)=(\partial y^{\mu}/\partial x^{\nu})v^{\nu}(x)$ do not change
the scaler product (2.2). Besides at a fixed spacetime point $x$ the tangent
vectors can be transformed by a second type of transformation along the
tangent vector axis which also preserve the scalar product (2.2):
\begin{equation}
v^{\prime a}=T_{b}^{a}(x)v^{b}, \tag{2.3}%
\end{equation}
where $T_{b}^{a}(x)$ are matrices satisfying the condition $\eta_{ab}T_{c}%
^{a}T_{d}^{b}=\eta_{cd}$. This means that the matrix elements $T_{b}^{a}(x)$
are elements of the group $SO(3,1)$ of special linear local transformations
with positive determinant depending on the spacetime point $x $ as a parameter.

Note that the scalar product (2.2) has also to be invariant with respect to
infinitesimal tangent vector line elements. This allows us to add translations
to the transformations in (2.3):%
\begin{equation}
v^{\prime a}=T_{b}^{a}(x)v^{b}+a^{a}(x), \tag{2.4}%
\end{equation}
and leads to the more general transformation group of a semi-direct product
$SO(3,1)\rtimes T(3,1)$.

The Poincare group and the transformation group (2.4) of tangent vectors in
the TB are described by the same group $SO(3,1)\rtimes T(3,1).$ However both
have a principal different geometrical and physical meaning: the first
transforms the coordinates of a flat spacetime manifold while the second
describes transformations within the tangent fiber $F=T_{x}(M)$ leaving the
spacetime point $x$ unchanged. In order to avoid confusions we denote
exclusively Poincare or Lorentz transformations as coordinate transformations
of the flat spacetime manifold. In mathematics the transformation group (2.4)
is denoted as a special affine group (SAG).

The TB is a special fiber bundle with the structure group $G=SO(3,1)\rtimes
T(3,1)$. In general, gauge theories with a gauge group $G$ can be described
using the mathematical formalism of fiber bundles with the group $G$ as
structure group. In this formalism gauge fields correspond to the connection
on the principle bundle while matter fields are described by representations
of the group $G$ on the associated vector bundle. A principle bundle $P(M)=$
($P,M,\pi,G)$ is a geometric structure over the base manifold $M$ (here the
spacetime manifold) with $G$ as the typical fiber, but $G$ act\ also as
structure (gauge) group on the fiber, $\pi$ is the projection from the bundle
$P(M)$ to the base manifold $M$. In quantum theory there is a fundamental
relationship between the irreducible representations of a symmetry group $G$
of a system and the space of quantum states of the system. Therefore group
representations must be included into the fiber bundle formalism. This is done
through the concept of vector bundles associated to the principle fiber
bundles. An associated fiber bundle $E(M,V,\pi,G,P)$ includes the vector space
$V$ into the bundle structure which is the vector representation of $G$ and
has the same structure group $G $ as the principle bundle $P(M)$.

Differential geometry on a fiber bundle can be executed by using the
definition of connections and covariant derivatives on the bundle. The
definition of a covariant derivative demands to consider vectors which point
from one fiber to the other at neighboring points $x$ and $x%
\acute{}%
$ of the spacetime manifold. The generators $\mathbf{L}_{a}$ of the group $G$
are vertical vectors pointing along the fibers and therefore belong to the
vertical subspace $V_{u}(P).$ Horizontal vectors in the subspace $H_{u}(P)$
point away from the fibers and are elements of the tangent space of the fiber
bundle $T_{u}(P)$ that complement the vertical vectors in $V_{u}(P)$. They can
be constructed by \cite{TG1,TG2}%
\begin{equation}
T_{u}(P)=H_{u}(P)\oplus V_{u}(P). \tag{2.5}%
\end{equation}
The complementary splitting of the tangent bundle into the vertical and
horizontal sub-bundle is an important concept in the theory of fiber bundles.
In the SM the vertical bundle describes the internal degree of freedom arising
by the gauge group. On the other hand the horizontal sub-bundle is a concept
to formulate the notion of a connection on the fiber bundle.\textbf{\ }

Let us now return to the specifics of the TB. The affine group
$G=SO(3,1)\rtimes T(3,1)$ is a semi-direct product and not semi-simple. As
explained in the introduction here it is assumed that gravity can be described
by the translational gauge theory with the structure group $T(3,1)$ based on
the translational connection as gauge field leading to the teleparallel
gravity theory \cite{t1,t2,g5,t3,t4,t5}. In contrast to this theory we assume
that the sub-group $SO(3,1)$ is related with the other fundamental
interactions of particle physics and the spin-connection do not vanish but is
related with the generalized electrowek interaction \cite{I1}. The generators
of the group $SO(3,1)$ are given by the 6 generators $\mathbf{M}_{ab}$ of the
group $SO(3,1)$ with $\mathbf{J}_{a}=\epsilon_{abc}\mathbf{M}_{bc}%
,\mathbf{K}_{a}=-\mathbf{M}_{oa}$ and the generators $\mathbf{P}$ $_{a}$ of
the translational group $T(3,1)$. This groups has two Casimir operators
$\mathbf{P}^{2}=\mathbf{P}^{a}\mathbf{P}_{a}$ and $\mathbf{w}^{2}%
=\mathbf{w}^{a}\mathbf{w}_{a}$ where the Pauli-Lubanski vector $\mathbf{w}%
_{a}$ is defined by%
\begin{equation}
\mathbf{w}_{a}=\frac{1}{2}\epsilon_{abcd}\mathbf{M}^{bc}\mathbf{P}^{d}
\tag{2.6}%
\end{equation}
The action of the affine group $G$ on a vector space $X$ is not transitive,
but the vector space decomposes into different orbits in which $G$ acts
transitively. A group action of $G$ on $X$ is transitive if for every pair of
elements $x$ and $y$, $x,y\in X$ there is a group element $g\in G$ such that
$gx=y$. The orbit of $X$ is the sub-manifold of X consisting of all points
that can be reached by acting on $X$ with $G$. The space $X$, which has a
transitive group action, is called a homogeneous space. Wigner developed the
method of induced representations to find the unitary representation of the
Poincare group. The Poincare group for spacetime transformations and the SAG
for tangent vector transformations are \ mathematical identical, therefore we
can use this method here. Since the translational generator $\mathbf{P}^{a}$
commute with each other the physical states can be expressed in term of the
eigenvalues of the action of $\mathbf{P}^{a}\Psi=k^{a}$ $\Psi.$ The subgroup
of $SO(3,1)$ with group elements $T_{b}^{a}(x)$ which leaves $k^{a}$ invariant
is determined by
\begin{equation}
T_{b}^{a}(x)k^{b}=k^{a} \tag{2.7}%
\end{equation}
and is called Wigner`s little group, $k^{a}$ is denoted as the standard
vector. There exist 6 orbits according to the eigenvalues of $\mathbf{P}^{2}$
and \textbf{P}$^{0}$. For $k^{2}\prec0$ with $k^{0}\succ0$ and $k^{0}\prec0$
the little group is $SO(3)$ with the generators $\mathbf{J}_{1},\mathbf{J}%
_{2},\mathbf{J}_{3}.$ For $k^{2}=0$ with $k^{0}\succ0$ and $k^{0}\prec0$ the
little group is the Euclidian group $E(2)$ with the generators $\mathbf{T}%
_{1}=\mathbf{K}_{1}+\mathbf{J}_{2},\mathbf{T}_{2}=\mathbf{K}_{2}%
-\mathbf{J}_{1}$ and $\mathbf{T}_{3}=\mathbf{J}_{3.}$ For $k^{2}\succ0$ the
little group is $SO(2,1).$

The description of matter fields in the quantum field theory (QFT) requires
the knowledge of the unitary representations $T_{L}(g)$ of these groups. The
composition law of the so-called vector representation $T_{L}(g)$ satisfy the
functional equation $T_{L}(g_{1})T_{L}(g_{2})=T_{L}(g_{1}g_{2})$ and encodes
the law of group transformations on the set of vector states. This leads to
certain pathologies, as for example the Dirac equation for massive particles
is not invariant under the Poincare group, but under its universal covering
group. Wigner solved this problem by using projective representations of the
Poincare group. In quantum theory the physical symmetry of a group of
transformations on a set of vector states has to preserve the transition
probability between two states $\left\vert \prec\Phi,T_{L}(g)\Psi
\succ\right\vert ^{2}=$ $\left\vert \prec\Phi,\Psi\succ\right\vert ^{2}$.
Therefore, generalized representations (denoted as projective representations)
are allowed which satisfy the more general composite law: \cite{r8,r9}
\begin{equation}
T_{L}(g_{1})T_{L}(g_{2})=\varepsilon(g_{1},g_{2})T_{L}(g_{1}g_{2}) \tag{2.8}%
\end{equation}
where $\varepsilon(g_{1},g_{2})$ is a complex-valued antisymmetric function of
the group elements with $\left\vert \varepsilon(g_{1},g_{2})\right\vert =1.$
Any projective representation of a Lie group $G$ is equivalent to the unitary
representation of the central extension of the group $G^{c}$. In general a
central extension $G^{c}$ of a group $G$ with elements $(g$,$\varsigma)\in$
$G^{c}$\ and $g\in$ $G,\varsigma\in$ $U(1)$ satisfy the group law
\begin{align}
(g,\varsigma)  &  =(g_{1},\varsigma_{1})\ast(g_{2},\varsigma_{2})\tag{2.9}\\
&  =((g_{1}\ast g_{2},\varsigma_{1}\varsigma_{2}\exp[i\xi(g_{1},g_{2}%
)],\nonumber
\end{align}
where $\xi(g_{1},g_{2})$ is the 2-cocycle satisfying the relation $\xi
(g_{1},g_{2})+\xi(g_{1}\ast g_{2},g_{3})=\xi(g_{1},g_{2}\ast g_{3})+\xi
(g_{2},g_{3}),\xi(e,g)=\xi(g,e).$ For the case of simply\ connected groups
like the rotation group $SO(3$) projective representations are obtained by
replacing the group $SO(3)$ by its universal cover $SU(2)$. However the
Euclidean group $E(2)$ is not semi-simple and the covering group
$\widetilde{E}(2)$ is not sufficient. One has to use a larger group: the
universal central extension $E^{c}(2).$\ This group includes in addition to
the group elements of $E(2)$ the group $U(1)$ of phase factors $\varepsilon
(g_{1},g_{2})$ with$\mid\varepsilon(g_{1},g_{2})\mid=1.$ The group $E^{c}(2)$
has been studied previously as e.g. in \cite{r9,r10,r11} and consists of
elements $(\alpha,\mathbf{a,}\omega)$ with $(\alpha,\mathbf{a})\in
E(2),\omega\mathbf{\in}R$.

The action of a group on a vector space determines the representation of the
group. The action of the group $E^{c}(2)$ on a vector $Z=(\xi_{1},\xi
_{2},\beta)$ is described by \cite{r9,r10,r11}.%

\begin{align}
(\alpha,\mathbf{a,\omega})(\xi_{1},\xi_{2},\beta)  &  =(\xi_{1}\cos
\alpha\mathbf{+}\xi_{2}\sin\alpha+a^{1},\tag{2.10}\\
&  -\xi_{1}\sin\alpha\mathbf{+}\xi_{2}\cos\alpha+a^{2},\nonumber\\
&  \beta+\omega+\frac{1}{2}m(\alpha,\mathbf{a,}\xi_{1},\xi_{2})),\nonumber
\end{align}
where $m(\alpha,\mathbf{a,}\xi_{1},\xi_{2})$ is the 2-cocycle which gives the
desired central extension parametrized as%
\begin{align}
m(\alpha,\mathbf{a,}\xi_{1},\xi_{2})  &  =(a_{1}\xi_{1}+a_{2}\xi_{2}%
)\sin\alpha-\nonumber\\
&  -(a_{1}\xi_{2}-a_{2}\xi_{1})\cos\alpha. \tag{2.11}%
\end{align}
\ \ From (2.10) and (2.11) the generators of the central extended $E^{c}(2)$
group can be calculated which are given by$:$%

\begin{align}
\mathbf{T}^{1}  &  =-i(\frac{\partial}{\partial\xi_{1}}+\frac{1}{2}\xi
_{2}\frac{\partial}{\partial\beta}),\tag{2.12}\\
\mathbf{T}^{2}  &  =-i(\frac{\partial}{\partial\xi_{2}}-\xi_{1}\frac{1}%
{2}\frac{\partial}{\partial\beta}),\nonumber\\
\mathbf{T}^{3}  &  =-i(\xi_{1}\frac{\partial}{\partial\xi_{2}}-\xi_{2}%
\frac{\partial}{\partial\xi_{1}}),\mathbf{E=-}i\frac{\partial}{\partial\beta
},\nonumber
\end{align}
which satisfy the following commutation rules $\mathbf{[T}^{1},\mathbf{T}%
^{2}]=i\mathbf{E,}[\mathbf{T}^{1},\mathbf{T}^{3}]=-i\mathbf{T}^{2},$
$[\mathbf{T}^{1},\mathbf{T}^{3}]=-i\mathbf{T}^{2},[\mathbf{T}^{1}%
,\mathbf{T}^{3}]=-i\mathbf{T}^{2},$ $[\mathbf{T}^{a},\mathbf{E}]=0.$

In general the (vertical) Laplacian (Casimir operator) of a gauge group plays
an important role in gauge theories, it delivers the description of the
"internal" space of elementary particles. Its eigenfunctions provide a
complete basis for the description of the basic states $\Phi_{vert}(u)$ of the
vertical subspace which are a part of the total wavefunction $\Phi
(x,u)=\Phi_{hor}\Phi_{vert}(u)$. The horizontal part $\Phi_{hor}$ depend on
the spacetime variable $x$ and is determined by the covariant derivative with
the connection on the associated bundle. The eigenvalue of the Laplacian
depend on the internal quantum numbers (IQN) such as hypercharge and iso-spin
in the SM. The vertical Laplacian of the group $E^{c}(2)$ is determined by%

\begin{equation}
\mathbf{\Delta}_{E^{c}}=(\mathbf{T}^{1})^{2}+(\mathbf{T}^{2})^{2}%
+2\mathbf{T}^{3}\mathbf{E.} \tag{2.13}%
\end{equation}
\emph{\ } Using polar coordinates $\xi_{1}=\xi\cos\phi,$ $\xi_{2}=\xi\sin\phi$
\ \ and $h_{nm\varkappa}(\xi,\beta,\phi)=\exp(i\varkappa\beta)(\exp
(im\phi)g_{nm\varkappa}(\xi)$ for the eigenfunctions of the Laplacian we find
with (2.12),(2.13) the following equation for the functions $g_{nm\varkappa
}(\xi):$\textbf{\ }
\begin{align}
&  [(-\frac{1}{\xi}\frac{\partial}{\partial\xi}\xi\frac{\partial}{\partial\xi
}+\frac{1}{\xi^{2}}m^{2})+\varkappa^{2}\xi^{2}-2\varkappa m]g_{nm\varkappa
}(\xi)\nonumber\\
&  =\epsilon_{nm\varkappa}g_{nm\varkappa}(\xi). \tag{2.14}%
\end{align}
The solutions of (2.14) for $h_{nm\varkappa}=h_{nm\varkappa}(\xi,\beta,\phi)$
are given by \
\begin{align}
h_{nm\varkappa}  &  =N_{nm\varkappa}\exp(i\varkappa\beta)\exp(im\phi
)g_{nm\varkappa}\tag{2.15}\\
g_{nm\varkappa}  &  =(\exp(-\frac{\mid\varkappa\mid\xi^{2}}{2})(\sqrt
{\varkappa}\xi)^{\mid m\mid}L_{n}^{\left\vert m\right\vert }(\mid\varkappa
\mid\xi^{2})\nonumber
\end{align}
with $N_{nm\varkappa}=\sqrt{\frac{\varkappa}{\pi}}(\frac{n!}{(\mid m\mid
+n)!})^{\frac{1}{2}}$, $\epsilon_{nm\varkappa}=4\varkappa(n+\frac{1}{2}%
+\frac{1}{2}(m+\mid m\mid),$ $n=0,1,2.$., $m=0,\pm1,\pm2,..,\varkappa=$%
,$\pm1,\pm2,..L_{n}^{\left\vert m\right\vert }(x)$ are the associated Legendre
polynomials and the IQN $m$ can be interpreted as hypercharge known from the
SM, but here two additional IQNs arise: the $E^{c}$-charge $\varkappa$ and the
family quantum number $n$ which could elucidate the existence of families in
the SM.

Note that the here considered problem of the central extension of the
Euclidian group $E(2)$ for symmetry transformations on the tangent bundle
differ from the analog problem for massless representations of the Poincare
group for spacetime transformations. Since massless particles are not observed
to have continuos degree of freedom, one put the requirement for the physical
states $\Psi_{k\sigma}$ \cite{r12}.%

\begin{equation}
\mathbf{T}^{1}\Psi_{k\sigma}=\mathbf{T}^{2}\Psi_{k\sigma}=0 \tag{2.16}%
\end{equation}
and physical states are only characterized by the operator $\mathbf{T}^{3}%
\Psi_{k\sigma}=\sigma\Psi_{k\sigma},$ where the operators $T^{i}$ are the
corresponding operators for massless representations of the Poincare group and
$\sigma$ is\ the helicity. Even though the Poincare group and the affine group
(2.4) are mathematical identical, there exist a significant difference in
physical applications.

The solutions $h_{nm\varkappa}$ in \ (2.15) form an orthonormal set and have
the analogous form like the solutions of the Schr\"{o}dinger equation in two
space dimensions for electrons in a perpendicular external magnetic field $B$
where the entity $\varkappa$ is substituted by $\varkappa\rightarrow eB/2$. In
the Schr\"{o}dinger equation, the different levels with quantum number
$n=0,1,2.$.. are denoted as Landau levels. Each Landau level is highly
degenerate, the degeneracy is $BA/2\pi$ where $A$ is the area of the system.
An important entity in the quantum Hall effect is the filling factor $\nu$ of
a Landau level defined as the ratio of the density of electrons $n_{el}$ to
the density of states in a Landau level $n_{\deg}=eB/2\pi$ (in the unit system
$c=1,h=2\pi,$ with $h$ as the Planck constant) given by $\nu=2\pi n_{el}/eB$.
The analogy of the solutions (2.15) with the solutions of the 2D
Schr\"{o}dinger equation in an external magnetic field suggests to define a TB
filling factor $\nu_{TB}$ as a characteristic entity. The variables $\xi
_{1},\xi_{2}$ are dimensionless variables of the representation of the group
$E^{c}(2)$. It is thus natural to define a dimensionless "density" $n_{TB}$ of
quarks in the vertical subspace. In analogy with the QHE the filling factor of
a TB-LL is given by
\begin{equation}
\nu_{TB}=\frac{\pi n_{TB}}{\mid\varkappa\mid}. \tag{2.17}%
\end{equation}

The generators $\mathbf{J}_{1},\mathbf{J}_{2},\mathbf{J}_{3\text{ }}$of the
semi-simple group $SU(2)$ \ are well known (see e.g.\cite{b1}) and, by using
the Laplacian $\Delta=(\mathbf{J}_{1}^{2}\mathbf{+J}_{2}^{2}\mathbf{+J}%
_{3}^{2})$ of this group all finite-dimensional eigenfunctions of the
Laplacian depending on the iso-spin IQN $I$ and $I_{3}$ can be found.

In the traditional quantum field theory, the internal degrees of freedom (such
as isospin or color) are described by spacetime- depend multi-component fields
taking into account the vertical structure given by the gauge groups by the
corresponding Lie-algebra representation. In the TB description, in contrast
to the multi-component formalism, the basic objects are functions $\Phi(x,u)$
depending both on the coordinate $x$ of the spacetime manifold and the
variables $u$ of the gauge group. For final dimensional representations of the
group $SU(2),$ complex coordinates $z^{1}$ and $z^{2}$ can be used for its
description. This is equivalent to the usual two-component description by
iso-spinors $(\phi_{1}(x),\phi_{2}(x))^{T}.$ However, for a non-semi-simple
group like the group $E^{c}(2),$ a multi-component representation is not
favorable, and the tangent bundle approach using coordinates of the tangent
space is more convenient. Therefore we use here in (2.12) and (2.13) the
generators of the group and not those of the Lie algebra as commonly done in
the SM.

The remaining little group $SU(1,1)$ is non-compact, and we do not consider
here whether it has a physical meaning as symmetry group for tangent fiber transformations.

\textbf{3}.\textbf{\ Emergent SU(3) symmetry and strong interaction in the
tangent bundle geometry}

\textbf{3.1 Analogy of the vertical TB Laplacian with the Hamiltonian of a
fractional quantum Hall system}

The Laplacian $\mathbf{\Delta}_{E^{c}}$ on the group $E^{c}(2)$ given by
(2.13) refer to a single tangent fiber at a fixed space point $\mathbf{x}%
$.\emph{\ }Now we consider the whole space with a finite volume $V=L^{3}$ and
present the fields arranges in regular cubic lattice of unit cells which are
defined by a set of position vectors $\mathbf{R}=n_{1}\mathbf{a}_{1}%
+n_{2}\mathbf{a}_{2}+n_{3}\mathbf{a}_{3}$ in which $n_{i}=$ $0,\pm1,\pm2$ run
over all integers and $\mathbf{a}_{i}$ are linear independent basis vectors.
The reciprocal lattice is defined by $G=l_{1}\mathbf{b}_{1}+l_{2}%
\mathbf{b}_{2}+l_{3}\mathbf{b}_{3}$. Using periodic boundary conditions, the
Fourier coefficients take discrete values $\mathbf{k}\rightarrow$
$\mathbf{k}_{i}=\frac{2\pi}{L}(l_{1i},l_{2i},l_{3i}),l_{i}=0,\pm1,\pm2...$In
the limit $L\rightarrow\infty$ we associate a tangent fiber with an elementary
cell with volume $\Delta V_{i}=(\frac{2\pi}{L})^{3}$. Due to the fermion
character of quarks and leptons each of the state in a lattice cell can be
occupied with one quark and one lepton of every possible internal quantum
number and helicity. The tangent bundle $TM$\ associates to every point $x\in
M$\ a vector space $T_{x}M$\ \ which is described by the variables of the
representation of the groups $E^{c}(2)$ and $SU(2)$. To obtain the vertical
Laplacian of the TB we combine the fibres attached at all space points $x_{i}%
$. For the $E^{c}(2)$ part of the vertical Laplacian of the bundle, we obtain
from (2.13)%

\begin{equation}
\Lambda_{E^{c}(2)}=%
{\displaystyle\sum\limits_{i}}
\mathbf{[}-i\frac{\partial}{\partial\xi_{1}^{(i)}}-\frac{\varkappa}{2}\xi
_{2}^{(i)}]^{2}+\mathbf{[}-i\frac{\partial}{\partial\xi_{2}^{(i)}}%
+\frac{\varkappa}{2}\xi_{1}^{(i)}]^{2} \tag{3.1}%
\end{equation}
where $\xi_{1}^{(i)},\xi_{2}^{(i)}$ are the corresponding variables of the
representation of the $E^{c}(2)$ group attached to the space cell $i$. The
Laplacian (3.1) has an analog form as the multi-particle Hamiltonian of a 2D
quantum mechanical many electron system in an external uniform magnetic field.
This Laplacian contains a rich internal structure, and describes a highly
correlated system characterized as a topological quantum liquid with closely
analogous to the integer and fractional quantum Hall effect.

The integer QHE\ has been discovered in 1980 by Klitzing et al. \cite{QH1} in
a 2D layer of a semiconductor at low temperature and strong magnetic fields.
In this experiment it was found that the Hall conductance takes quantized
values of $\sigma_{xy}=e^{2}\nu/2\pi\hbar$ , where $\nu$ is precisely an
integer number, $\nu=1,2$,.... The fractional QHE was discovered by Tsui,
Stormer and Gossard in 1982 \cite{QH2}, who reported that $\nu$ is not only
restricted to take integer values, but can take values at $\nu=1/3$ and
$\nu=2/3.$

The integer QHE can be understood because the 2D electron gas forms an
incompressible liquid at the filling factors $\nu=n=1,2,3$ due to the Landau
level structure with a finite energy gap for all charged excitations. The
fractional QHE was first explained by a theory of Laughlin \cite{QH3}. He
proposed a trial ground-state many-body wavefunction in a partially filled
Landau level with filling fraction $\nu=1/(2p+1),p=1,2,3..$which include
strong Coulomb interaction and correlations among the electrons.

Even though the issue of the QHE is now well understood and described in
excellent reviews and books (see e.g. \cite{H1,H4,H5} )we choose a
self-contained presentation accessible to readers who are not specialized in
solid state theory.

In the TB-QFT the ground state (vacuum state) with the lowest possible energy
is described by the family IQN $n=0.$\ In traditional QFT the vacuum state is
a Fock-state which contains no physical particles but quantum fluctuations of
virtual particles related with the non-commutation of the quantized fields. In
this case the vacuum expectation value of any field operator vanishes; in
particular, the particle density of the vacuum is zero. The close analogy to a
QH system requires a redefinition of the vacuum in the TB-QFT. As explained in
more detail in chapter 4 the analog form of the Laplacian with that of a
quantum Hall system can be interpreted by the hypothesis that the vacuum with
the IQN $n=0$ is filled with see leptons and composite see quarks and all
higher levels $n=1,2,3$ are empty. Furthermore, there is an energy gap between
the lowest TB-LL $n=0$ and a higher TB-LL. If a seequark is excited into a
higher TB-LL, it leaves a hole in the old state. A quark hole carry the
opposite hypercharge and opposite isospin IQN, but a positive energy, and can
be interpreted as an anti-quark. There exists also several other physical well
established facts in QCD as the existence of quark condensation, Higgs
condensation and others which hint on a non trivial vacuum structure similar,
as a condensate in solid state theory.

Let us first study the eigenvalue equation of the vertical Laplacian for the
$E^{c}$ group of the TB%

\begin{equation}
\Lambda_{E^{c}}\Phi_{vert}=\varepsilon\Phi_{vert} \tag{3.2}%
\end{equation}
where $\Phi_{vert}$ is the vertical part of the total wavefunction $\Phi
=\Phi(x)\Phi_{vert}$ describing the internal degree of freedom of the group
$SU(2)\otimes$\textbf{\ }$E^{c}(2).$ \textbf{\ }To simplify the analysis we
first ignore the iso-spin degree of freedom and assume that all iso-spins are
polarized and frozen into one direction. For the completely filled ground
state $n=0$ the wave function of $N$ fermions can be described by a Slater
determinant. In every cell $i$ particles with different hypercharge numbers
$m$ are placed which run from $0$ to $N-1$. Using the single-particle wave
function (apart from a normalization constant) and variables $\eta_{i}=\xi
_{1}^{(i)}+i\xi_{2}^{(i)}=\xi_{i}\exp(i\varphi_{i})$ the many-particle
solution can be determined from (3.2)\ by
\ \ \ \ \ \ \ \ \ \ \ \ \ \ \ \ \ \ \ \ \ \ \ \ \ \ \ \ \ \ \ \ \ \ \ \ \
\begin{align}
\Phi_{vert}^{0}(u)  &  =S\exp(i\varkappa\beta)\exp[-\frac{1}{4l_{\varkappa
}^{2}}%
{\displaystyle\sum\limits_{r=1}^{N}}
\mid\eta_{r}\mid^{2}]\tag{3.3}\\
&  =%
{\displaystyle\prod\limits_{r\prec q}^{N-1}}
(\eta_{r}-\eta_{q})\exp(i\varkappa\beta)\exp(-\frac{1}{4l_{\varkappa}^{2}}%
{\displaystyle\sum\limits_{r=1}^{N}}
\mid\eta_{r}\mid^{2})\nonumber
\end{align}
with the Slater determinant $S\ $\ as
\[
S=\det%
\begin{bmatrix}
1 & 1 & 1 & ...\\
\eta_{1} & \eta_{2} & \eta_{3} & ...\\
\eta_{1}^{2} & \eta_{2}^{2} & \eta_{3}^{2} & ...\\
. & . & . & ...\\
\eta_{1}^{N-1} & \eta_{2}^{N-1} & \eta_{3}^{N-1} & ...
\end{bmatrix}
\]
and $l_{\varkappa}^{2}=\mid\varkappa\mid/2$. Since the highest power of any
particle coordinate $\eta_{r}$ is $N-1$ the maximum eigenvalue m is $m_{\max
}=N-1.$ The solution describes a fully occupied lowest TB-LL with a filling
fraction $\nu_{TB}=1.$ The wavefunction vanishes whenever two fermions come
together and the exponential factor decreases quickly with increasing
$\mid\eta_{r}\mid^{2}.$When we add one fermion state to the completely filled
system it is placed into a higher TB-LL level $n=1,2,$ .. because of the Pauli
exclusion principle.

In a 2D semiconductor in an external uniform magnetic field the analog
solution (3.3) describes a remarkable macroscopic quantum phenomenon: the
integer quantum Hall effect which is explained by a fixed and well-defined
density of the system with a filling fraction $\nu$\ given by $\nu=n$\ ,
$n=1,2,$...One can use (3.3) as the solution for the ground state $n=0$ of the
vertical part of the wavefunction of leptons.

Laughlin made a brilliant ansatz for a trial fermion wave function (known as
the Laughlin wave function) that describes the fractional quantum Hall effect
\cite{QH3}. Since the vertical Laplacian (3.1) has the same form as the 2D
multi-particle Hamiltonian in a quantum Hall system we can use the analog
Laughlin wavefunction $\Phi_{vert}^{L}(\eta)$ for the description of the
vertical part of the quark wavefunction in the vacuum $n=0$:
\begin{equation}
\Phi_{vert}^{L}(\eta)=%
{\displaystyle\prod\limits_{i\prec j}}
(\eta_{i}-\eta_{j})^{k}\exp(-\frac{1}{4l_{\varkappa}^{2}}%
{\displaystyle\sum\limits_{i}}
\mid\eta_{i}\mid^{2}) \tag{3.4}%
\end{equation}
where $k=2p+1$ must be an odd integer for $\Phi_{vert}^{L}(\eta)$ to be
totally anti-symmetric$.$This wavefunction describes a uniform "density" state
with a partially filled lowest TB-LL with filling factor $\nu_{TB}=1/(2p+1)$.
Due to a particle-hole symmetry there exist also states at the filling factor
$\nu_{TB}=1-1/(2p+1)$ \cite{QH3}. In particular for $p=1$\ we find states with
the filling factor $\nu_{TB}=1/3$\ and $\nu_{TB}=2/3.$

Laughlin achieved a physical understanding of the wavefunction (3.4) by an
analogy with a 2D plasma. The joint probability distribution $P=\mid
\Phi_{vert}^{L}\mid^{2}=\exp(-\beta U(\eta_{1}...\eta_{N}))$ can be described
by a potential $U$. By using (3.4) and choosing $\beta=2/k$ the potential $U$
\ becomes $\ $
\begin{equation}
U=-k^{2}%
{\displaystyle\sum\limits_{i\prec j}}
\ln(\frac{1}{l_{\varkappa}}(\mid\eta_{i}-\eta_{j}\mid)+\frac{k}{4l_{\varkappa
}^{2}}%
{\displaystyle\sum\limits_{i}}
\mid\eta_{i}\mid^{2} \tag{3.5}%
\end{equation}
Equ.(3.5)\ has the form of a Coulomb potential of electrons in a $2D$
one-component plasma. The first term has the analog form as the Coulomb
interaction term between charged particles in $2D$ characterized by a
logarithmic potential with a charge $q=-k.$ The second term describes their
interaction with a neutralizing background of positive charges similar as in
the yellium model. The analog for the "density" of the background charge is
$\rho_{B}=1/2\pi l_{\varkappa}^{2}=\mid\varkappa\mid/\pi$ (note that $\rho
_{B}$ describes a dimensionless entity in the $E^{c}$ group manifold). The
complete screening property of the hypercharge in the ground state now
requires, similar as in a plasma, that the "charge" density of the plasma
particles of see quarks in the ground state is equal to the background charge
density $\rho_{B}=$ $\mid\varkappa\mid/\pi.$ Each particle carry the charge
$q=-k$, thefore the compensating "density" of particles in the ground state
$n=0$ is $k\varrho_{vac}=\rho_{B},$ or
\begin{equation}
\varrho_{vac}=\frac{\mid\varkappa\mid}{\pi(2p+1)} \tag{3.6}%
\end{equation}
This is the "density" of a state at filling fraction
\begin{equation}
\nu=\frac{1}{(2p+1)} \tag{3.7}%
\end{equation}

The effect of adding or removing one particle from the ground state
corresponds to the creation of a quasi-particle or a quasi-hole. A quasi-hole
located at $\eta_{0\text{ }}$can be described by the wavefunction generated by
acting on the vacuum state as follows \cite{QH3}]%
\begin{equation}
\Phi_{vert}^{h}(\eta)=%
{\displaystyle\prod\limits_{i=1}^{N}}
(\eta_{i}-\eta_{0})\Phi_{vert}^{L}(\eta) \tag{3.8}%
\end{equation}
obtained from Laughlins wave function $\Phi_{vert}^{L}(\eta)$ by introducing
the first factor $(\eta_{i}-\eta_{0}).$ It vanishes at the position
corresponding to the complex parameter $\eta_{0}.$This means it describes a
state where the quark "density" is zero at $\eta_{0}$ and therefore
characterizes a hole. It is an excited state because it has an additional
vortices and the hypercharge is boosted by one unit. To calculate the
quasihole charge one can use the plasma analogy with the wave function
(3.8).The joint probability distribution of the quasihole is given by
\begin{equation}
P=\mid\Phi_{vert}^{h}(\eta)\mid^{2}=\exp\{-\frac{1}{2k}[(U-k\ln(\frac
{1}{l_{\varkappa}}\mid\eta_{i}-\eta_{0}\mid]\} \tag{3.9}%
\end{equation}
where $U=U(\eta_{1}...\eta_{N})$ is given in (3.5).\ Since a system like a
plasma attempts to maintain charge neutrality according to (3.9) it will
screen the hole by the charge of $k$ plasma particles. The resulting screening
cloud has a net deficit of $1/k$ vacuum particles. This means that the
quasihole represents an excitation with a fractional charge $e_{h}%
=e/k=e/(2p+1).$The above described analogy with the fractional QHE here arise
in a natural way from the bundle Laplacian (3.1) of the group $E^{c}(2)$ and
can explain charge quantization of leptons with $m_{lep}=1$ and fractional
charges of quarks with a quasi-hole hypercharge given by $m_{q}=1/(2p+1.$The
fractional charges of quasi-holes and quasi-particles can also be proven by
other methods (see e.g. \cite{H4}).

The hypercharge of an excited valence quark should carry the opposite sign as
in (3.8). The corresponding wavefunction should contain an antivortex which
involves $\eta_{j}^{\ast}$, but then the wavefunction do not sit in the lowest
TB-LL, and rather one has to project it into it. According this procedure the
wavefunction proposed by Laughlin for the quasiparticle state can be used for
excited valence quarks as \cite{QH3}
\begin{equation}
\Phi_{vert}^{q}=%
{\displaystyle\prod\limits_{i=1}^{N}}
(2l_{\varkappa}^{2}\frac{\partial}{\partial\eta_{i}}-\eta_{0}^{\ast}%
)\ \Phi_{vert}^{L}(\eta) \tag{3.10}%
\end{equation}
where the derivatives act only on the polynomial part of $\Phi_{vert}^{L}%
(\eta).$

Another important feature of the wavefunction \ is the existence of
vortices-here we denote it as TB vortices. A TB vortex is a winding in the the
phase of the variable $\eta_{r}=\xi_{1}^{r}+i\xi_{2}^{r}=\xi_{r}\exp
(i\varphi_{r}).$ This means the phase $\varphi_{r}$ changes by $2\pi$ if the
particle moves around $\eta_{r},$

In the discussion above it has been assumed that in the vacuum, all states
carry the same isospin number. Generalization of this situation is possible to
systems where the quarks are not isospin polarized but the lowest TB-LL is
occupied by both isospin states, forming an isospin singlet. The simplest
two-component wave function describing a topological state filled with
particles with spin in both directions was given by Halperin \cite{QH4}. Here
we use it for the description of the vertical part of wavefunctions for
isospin singlet states of quarks with two iso-spin components. According
\cite{QH4} this solution has the form
\begin{align}
\Phi_{vert}^{H}  &  =%
{\displaystyle\prod\limits_{i\prec j}^{N_{\uparrow}}}
\exp[-\frac{1}{4l_{\varkappa}^{2}}(%
{\displaystyle\sum\limits_{i}}
(\mid\eta_{i}^{\downarrow}\mid^{2}+\mid\eta_{i}^{\uparrow}\mid^{2}%
)]\tag{3.11}\\
&  \ \times(\eta_{i}^{\uparrow}-\eta_{j}^{\uparrow})^{p+1}(\eta_{i}%
^{\downarrow}-\eta_{j}^{\downarrow})^{p+1}(\eta_{i}^{\uparrow}-\eta
_{j}^{\downarrow})^{p}\nonumber
\end{align}
\textit{\ }where the set of coordinates $\eta_{i}^{\uparrow},i=1...N$
$\ $corresponds to iso-spin up and $\eta_{i}^{\downarrow}$, $i=1...N$ to
iso-spin down quarks, $p$ is an integer number. The filling factors are given
by
\begin{equation}
\nu_{\uparrow}=\nu_{\downarrow}=\frac{1}{1+2p}. \tag{3.12}%
\end{equation}

Note that in experimental studies of the anomalous QHE also filling factors
with other rational numbers than (3.7) or (3.12) has been observed and several
theoretical explanations has been developed.\textit{\ }

\textbf{3.2} \textbf{Composite quarks with fractional hypercharges}

The composite-fermion picture \cite{H5, Q5} is a convenient way to provide an
intuitive idea for the fractionally charged quasi-particles and various other
aspects of the quantum Hall effect. This concept can be transferred to the
understanding of fractional charged quarks in the TB geometry. When all
particles are confined in the lowest TB-LL the wave function is a polynomial
of the complex variable $\eta=\xi_{1}+i\xi_{2}=\xi\exp(i\phi)$ which has a
vortex at the origin because a complete loop around the origin changes $\phi$
by $2\pi.$ A composite fermion is simply envisioned as a bound state of a
fermion carrying an even number of quantized vortices of the many particle
wave function. This can be understood as a screening of the fermions, since
vortices create holes around each fermion. Composite quarks capture two TB
vortices and as a consequence experience a much reduced effective interaction.
Thus in this approach strongly correlated quarks are mapped into weakly
interacting quarks with an effective reduced IQN $\varkappa_{eff}$. The bound
state of a quark and two vortices can be described qualitatively by a simple
reformulation of the Laughlin function (3.4) by \cite{Q5}
\begin{equation}
\Phi_{vert}^{J}=%
{\displaystyle\prod\limits_{i\prec j}}
(\eta_{i}-\eta_{j})^{2}\ \Phi_{vert}^{L}(\eta) \tag{3.13}%
\end{equation}
Here $\Phi_{vert}^{L}(\eta)$ is the Laughlin state given by (3.4)\ and the
first factor represents two TB vortices that are attached to each coordinate
$\eta_{i}$ where a quark is present. Thus the Laughlin ground state
$\Phi_{vert}^{L}(\eta)$ for partially filled quarks of the state $n=0$ is now
understood as a Jain ground state $\Phi_{vert}^{J}$ completely filled with
composite quarks.

A composite quark experience a reduced IQN $\varkappa_{eff}$. This can be
calculated in an analogous way as in the case of the anomalous QHE in
semiconductors taking into account the magnetic field of the $2p$ attached
vortices $B_{vort}=-2p\rho_{0}\Phi_{0}$ pointing antiparallel to the external
magnetic field B, where $\Phi_{0}=h/e$ is the elementary quantum of magnetic
flux. Thus with the effective magnetic field $B_{eff}=B-2p\rho_{0}\Phi_{0}$
the composite fermions experience a reduced effective magnetic field. In
\ analogy, the fractional filling factor of bare quarks \ $\nu_{TB}=1/(2p+1)$
gives for the composite quarks a completely filled ground state $\nu_{eff}=1.$
Using the analogy of $\varkappa$ with the magnetic field $B,$ we can
substitute $eB_{eff}\rightarrow2\varkappa_{eff}$ \ and with $\rho_{0}=\nu/\pi
$, $\Phi_{0}=2\pi$ $\ ($for $c=1,h=2\pi$) and $\nu_{TB}=1/(2p+1)$\ we obtain%

\begin{equation}
\varkappa_{eff}=\varkappa-\varkappa\frac{2p}{2p+1}=\varkappa\frac{1}{2p+1}.
\tag{3.14}%
\end{equation}

This means that in TB-QFT not only the effective hypercharge $m_{eff}%
=(2p+1)^{-1}$ receive a fractional number, but also the $E^{c}$ charge
$\varkappa_{eff}.$

\textbf{3.3 Quark and lepton families, fractional hypercharges and fractional
}$\mathbf{E}^{c}$\textbf{\ charges of quarks}

In the SM the members arranged in different families of leptons and quarks
have identical IQNs and properties except for their masses. In contrast in the
TB-QFT different families can be distinguished by the different family quantum
numbers $n=1,2,3,$ while the vacuum state carries the TB family quantum number
$n=0$.

In the first family with $n=1$ there are an iso-spin doublet ($I_{3}=\pm1/2$)
\ with left- handed electrons and electron-neutrinos $(e_{L,}\nu_{e}{}_{L}),$
an iso- spin singlet ($I_{3}=0)$ with right-handed electrons $e_{R}$ , and
also corresponding up and down quarks $u$ and $d:$%
\[
\binom{\nu_{e}}{e}_{L},e_{R};\binom{u}{d}_{L},u_{R},d_{R}.
\]
In the second family with $n=2$ we have the muon $\mu$ and its neutrino as
well as charm (c) and strange (s) quarks:%
\[
\binom{\nu_{\mu}}{\mu}_{L},\mu_{R};\binom{c}{s}_{L},c_{R},s_{R}.
\]
In the third family $n=3$ we have the tauon $\tau$ and its neutrino, as well
as top (t) and bottom (b) quarks%
\[
\binom{\nu_{\tau}}{\tau}_{L},\tau_{R};\binom{t}{b}_{L},t_{R},b_{R}.
\]
In TB-QFT, in addition to hypercharge and isospin, the new family IQN $n$ and
the E$^{c}$ charge $\varkappa$ exist. Rules for the determination of the new
IQN $\varkappa$ can be found by the Yukawa interaction Lagrangian, which we
write in a compact form by using the notation $Q_{n}=(U_{n,}D_{n})_{L}%
^{T},(U_{n})_{R},(D_{n})_{R}$, where $n=1,2,3$ stands for the different
families.The Yukawa interaction Lagrangian can be written as
\begin{equation}
\mathcal{L}_{Yuk}=-%
{\displaystyle\sum\limits_{i,j}}
[c_{ij}^{D}\overline{Q_{i}}\Phi_{H}(D_{j})_{R}+c_{ij}^{U}\overline{Q_{i}}%
\Phi_{H}^{c}(U_{j})_{R}+h.c. \tag{3.15}%
\end{equation}
with $c_{ij}^{D}$ and $c_{ij}^{U}$ as constant coupling coefficients and
$\Phi_{H}^{c}=i\sigma_{2}\Phi_{H}.$ Inserting the expansion (2.15) into
$L_{Yuk}$ the Yukawa interaction Hamiltonian is easy to build with analog
expressions as in the SM but including additional matrix elements
\begin{equation}
I_{QHU_{R}}^{Y}=\int d\mu\chi_{H}(u)\chi_{Q}^{\ast}(u)\chi_{U_{R}}(u),
\tag{3.16a}%
\end{equation}
and%
\begin{equation}
I_{QH^{c}D_{R}}^{Y}=\int d\mu\chi_{H^{c}}(u)\chi_{Q}^{\ast}(u)\chi_{D_{R}}(u),
\tag{3.16b}%
\end{equation}
with the integration measure $d\mu(u)$ $=d\mu_{SU(2)}d\mu_{E^{c}}.$ From
(3.16) we see that the Yukawa interaction is nonzero only if in addition to
leptons and quarks the Higgs particle also carry a nonzero $E^{c}$ charge
$\varkappa_{H}.$ The matrix elements are non-zero if the following selection
rules for the hypercharges $m_{eff}$ of quarks are fulfilled%

\begin{align}
-m_{Q}+m_{H}+m_{D_{R}}  &  =0\tag{3.17}\\
-m_{Q}+m_{H^{c}}+m_{U_{R}}  &  =0\nonumber
\end{align}
where $m_{H}=-m_{H^{c}}=1.$Besides we have also the analog relations for the
$E^{c}$ charge $\varkappa:$
\begin{align}
-\varkappa_{Q}+\varkappa_{H}+\varkappa_{D_{R}}  &  =0\tag{3.18}\\
-\varkappa_{Q}+\varkappa_{H^{c}}+\varkappa_{U_{R}}  &  =0\nonumber
\end{align}
The solutions of the equations for the unknown $E^{c}$ charges $\varkappa$ are
not unique. Since interaction processes favor fractions with small
denominators, a possible special solution of the above equation system is
given by
\begin{equation}
m_{eff}=\varkappa_{eff}. \tag{3.19}%
\end{equation}
This means fractional $E^{c}$ charges $\varkappa$ of quarks are required by
the Yukawa interaction. Note that this results agrees with the composite
fermion picture discussed in caption 3.2\ .

Let us summarize the extension of the IQNs in the TB-QFT and its
interpretation by the composite quark picture. For the left-handed quarks
($D_{L})$ we have the IQNs $M_{D_{L}}=\{m_{eff}=\frac{1}{3},\varkappa
_{eff}=\frac{1}{3},I_{3}=-\frac{1}{2}\}$ and $M_{U_{L}}=\{m_{eff}=\frac{1}%
{3},\varkappa_{eff}=\frac{1}{3},I_{3}=\frac{1}{2}\}.$ For the right-handed
down-quarks ($D_{R})$ we have $M_{_{D_{R}}}=\{m_{eff}=-\frac{2}{3}%
,\varkappa_{eff}=-\frac{2}{3},I_{3}=0\}$ and for the right-handed up-quark
states $M_{_{U_{R}}}=\{m_{eff}=\frac{4}{3},\varkappa_{eff}=\frac{4}{3}%
,I_{3}=0\}.$ \ On the other hand the IQN of the Higgs are $M_{H}%
=\{n_{H},m=1,\varkappa=1,I_{3}=-\frac{1}{2}\}.$

Now one can explain the IQNs of different quarks according its composition by
bare quarks and attached TB vortices. As noted above, left-handed quarks
($U_{L,}D_{L})$ have the IQN $m_{eff}=1/3$ and $\varkappa_{eff}=1/3$ which can
be interpreted as excitations of composite holes with two attached TB
vortices. Right-handed $D_{R}$ quarks are isospin singles wih $I_{3}=0$ and
can be interpreted by the assumption that the lowest TB-LL is occupied\ with
isospin up and isospin down quarks with total filling fraction $\nu_{eff}=2$.
Since every isospin component carries the hypercharge $-1/3$ the hypercharge
of the $D_{R}$ quark is $m_{eff}=-2/3,$ $\varkappa_{eff}=-2/3.$ The
right-handed $U_{R\text{ }}$quarks with $I_{3}=0$ can be identified as excited
hole states with two isospin\textbf{\ }components, each of which carries the
hypercharge $m_{eff}=$ $2/3$. This means in agreement with the above remark
the total hypercharge of the $U_{R\text{ }}$quarks is $m_{eff}=4/3$ and of the
E$^{c}$ charge is $\varkappa_{eff}=4/3.$

The left-handed leptons ($N_{,}E)_{L}$ are interpreted as fermions without
attached TB vortices with $m=-1$ and $\varkappa=-1$ and a completely filled
lowest TB-LL $\nu=1.$ Right-handed leptons $E_{R}$ are isospin singlets with
$m=-2$ and $\varkappa=-2$ and filling factor $\nu=2.$

\textbf{3.4. Emergent Chern-Simon U(1) gauge fields on the tangent bundle }

In the composite fermion picture of the QHE \cite{H5, Q5}, the Laughlin
wavefunction (3.4) can be understood as a binding of an even number of
vortices to each fermion, turning it into a composite fermion. Later, a
field-theoretical formalism \ was developed, that allows a simple
understanding of the composite fermion concept and systematic calculations of
measurable quantities. In this "Chern Simon" approach, a singular gauge
transformation is used to map the Hamiltonian of interacting electrons to one
of electrons coupled to an additional emergent gauge field. Thus, the
fractional quantum Hall effect has a hidden dynamically generated emergent
local $U(1)$ gauge symmetry which is responsible for the binding of vortices
to the fermions. In this way, the Laughlin state with partially filled lowest
Landau level is mapped into a completely filled fermionic state of composite
fermions \cite{H3}. The Chern-Simon approach was first introduced for the
description of a boson superfluid interacting with an emergent gauge field
\cite{H2}.

Here the field-theoretical description of the anomalous quantum Hall effect is
used for the analog problem of the existence of emergent symmetries in the
Laplacian of the group $E^{c}(2)$. Let us first analyze the vertical part of
the wavefunction of the TB determined by the bundle Laplacian $\Lambda_{E^{c}%
}$ as given in (3.1). In the Chern-Simon approach a singular gauge
transformation of the Laplacian (3.1) is considered given by (see
e.g.\cite{Q5})\
\begin{equation}
\Psi^{`}(\mathbf{\xi}_{1},..,\mathbf{\xi}_{N})=U\Psi(\mathbf{\xi}%
_{1},...\mathbf{\xi}_{N}) \tag{3.20}%
\end{equation}
with%
\[
U=%
{\displaystyle\prod\limits_{j\succ k}}
\exp(-ik\theta(\mathbf{\xi}_{j}-\mathbf{\xi}_{k}))
\]
with $\mathbf{\xi}_{i}=(\xi_{1}^{i},\xi_{2}^{i})$ and $\theta(\mathbf{\xi}%
_{j}-\mathbf{\xi}_{k})=\operatorname{Im}\ln(\mathbf{\xi}_{j}-\mathbf{\xi}%
_{k})$. An even number $k=2p$ is required for a transformation from the bare
fermions to composite fermions. Performing \ the unitary transformation
$U^{-1}\Delta_{E^{c}}U$ \ the transformed Laplacian of (3.1) can be written as%

\begin{align}
\Delta_{E^{c}}^{%
\acute{}%
}  &  =%
{\displaystyle\sum\limits_{i}}
\mathbf{[}-i\frac{\partial}{\partial\xi_{1}^{i}}+\frac{\varkappa}{2}\xi
_{2}^{i}-A_{1}(\mathbf{\xi}^{i})]^{2}\tag{3.21}\\
&  +\mathbf{[}-i\frac{\partial}{\partial\xi_{2}^{i}}-\frac{\varkappa}{2}%
\xi_{1}^{i}-A_{2}(\mathbf{\xi}^{i})]^{2}\nonumber
\end{align}
where $A_{a}\mathbf{(\xi}^{i})$ (with $a=1,2)$ is an auxiliary (emergent)
field arising by the unitary transformation (3.20) and defined by
\begin{equation}
A_{a}\mathbf{\ (\xi}^{i})=k%
{\displaystyle\sum\limits_{j\neq i}}
\frac{\partial}{\partial\xi_{a}^{i}}\theta(\mathbf{\xi}_{j}-\mathbf{\xi}_{i}).
\tag{3.22}%
\end{equation}
Therefore we find
\begin{equation}%
{\displaystyle\oint\limits_{i}}
A_{a}\mathbf{(\xi}^{i})d\xi_{a}^{i}=2\pi k \tag{3.23}%
\end{equation}
where the integral is around any closed contour surrounding only the particle
$i$. The two-component vector potential $A_{a}(\mathbf{\xi})$ is related with
an emergent field $B(\mathbf{\xi)}$ (which is the analog of the "magnetic
field strength" in the 2-dimensional $E^{c}$ manifold)
\begin{align}
B(\mathbf{\xi)}  &  =\varepsilon^{ab}\partial_{a}A_{b}(\mathbf{\xi}%
)\tag{3.24}\\
&  =2\pi k\varrho(\mathbf{\xi}).\nonumber
\end{align}
where $\rho(\mathbf{\xi})=\sum_{k}\delta^{2}(\mathbf{\xi}-\mathbf{\xi}_{k})$
is the local "particle density" in the $E^{c}(2)$ manifold\textbf{\ }and
$\varepsilon^{ab}$ is the antisymmetric 2D Levi-Civita symbol.

The Laplacian (3.21) can be written in field-theoretical formalism%

\begin{align}
\Lambda_{E^{c}}^{%
\acute{}%
}  &  =\Phi_{vert}^{\dagger}(\mathbf{\xi})\{[-i\frac{\partial}{\partial\xi
^{1}}+\varkappa\frac{1}{2}\xi^{2}-A_{1}(\mathbf{\xi})]^{2}\tag{3.25}\\
&  +\mathbf{[}-i\frac{\partial}{\partial\xi^{2}}-\varkappa\frac{1}{2}\xi
^{1}-A_{2}(\mathbf{\xi})]^{2}\}\Phi_{vert}(\mathbf{\xi})\nonumber
\end{align}
where $\Phi_{vert}(\mathbf{\xi})$ is the $E^{c}(2)$ part of the vertical quark
wavefunction. The "density" $\rho_{vert}(\mathbf{\xi})$ in the
field-theoretical formalism is given by%
\begin{equation}
\varrho_{vert}(\mathbf{\xi})=\Phi_{vert}^{\dagger}(\mathbf{\xi})\Phi
_{vert}(\mathbf{\xi}). \tag{3.26}%
\end{equation}
We obtain a dynamic description if we introduce the time variable $\xi_{0}=t$
together with $\xi_{1}$ and $\xi_{2}$ and a new time component $A_{0}$ of the
emergent potential $A_{a}$. One can use the gauge condition $\partial^{a}%
A_{a}=0$ $(a=0,1,2).$ If we take the time derivative of (3.24) and use the
equation of conserved current $\partial_{t}\varrho+\partial_{a}j_{a}=0$ one
obtain
\begin{equation}
\varepsilon^{ab}\partial_{t}\partial_{a}A_{b}(\mathbf{\xi})=2\pi k\partial
_{t}\varrho(\mathbf{\xi}) \tag{3.27}%
\end{equation}
and consequently
\begin{equation}
\varepsilon^{ab}\partial_{t}A_{b}=-2\pi kj^{a}. \tag{3.28}%
\end{equation}

The relation \ (3.24) is a constrain for the "density" $\varrho(\mathbf{\xi})$
which is proportional to the emergent "magnetic field" $B(\mathbf{\xi).}$ In
the theory of the QHE the analog constrain is denoted as flux attachment to
the electrons. This relationship is a crucial feature in the QHE as well as in
the here studied TB-QFT. The second constraint (3.28) ensures that the
charge-flux relation is preserved under time evolution, because the time
derivation of (3.24) implies (3.28).

\ Alternatively the constrains (3.24) and (3.28) can also be taken into
account if we add an additional term to the Laplacian (3.25) (denoted as the
Chern-Simon action)
\begin{align}
L_{CS}  &  =\Phi_{vert}^{\dagger}(\mathbf{\xi})[i\frac{\partial}{\partial
\xi^{0}}-A_{0}(\mathbf{\xi})]\Phi_{vert}(\mathbf{\xi})+\nonumber\\
&  +\Lambda_{E^{c}}^{%
\acute{}%
}+\frac{1}{4\pi k}\varepsilon^{abc}A_{a}(\mathbf{\xi)}\partial_{b}%
A_{c}(\mathbf{\xi)} \tag{3.29}%
\end{align}
(with $\Lambda_{E^{c}}^{%
\acute{}%
}$ given in $(3.25),a,b=0,1,2$).The Chern-Simon action is described by the
last term in (3.29). In $D=2+1$ dimensions a particle current coupled to a
Chern-Simon field produces states with fractional charges and the binding of
particles to fluxes. This can be seen by the variation of (3.29) with respect
of $A_{0}$ yielding (3.24) and with respect to $A_{a}$ ($a=1,2$) which
delivers
\begin{equation}
\varepsilon^{ab}(\partial_{b}A_{0}-\partial_{0}A_{b})=2\pi kj^{b}(\mathbf{\xi
}). \tag{3.30}%
\end{equation}
(3.30) agrees with (3.28) using the special gauge $A_{0}=0.$ The coupling
constant $k$ has to be an even number: $k=2p$.

The Chern-Simons theory is in $(2+1)D$ \ a \ new type of gauge theory,
completely different from Maxwell theory in $(2+1)D.$ This theory is
particular interesting for its practical application in the QHE and describes
a deep analogy with quantum mechanical Landau levels and fractional charge
quantization. The effect of the Chern-Simon term in (3.29) is to tie the
magnetic field $B(\mathbf{\xi)}$ to the number density as described in (3.24).

A simple way to analyze the Chern-Simon approach is to make the mean field
approximation in the gauge $A_{0}=0.$ Using an average over the variables
$\mathbf{\xi}$ of the $E^{c}$ manifold $\prec\rho(\mathbf{\xi})\succ=\rho_{0}$
according to (3.24) the TB Chern-Simons field is smeared out to $\prec
B(\mathbf{\xi})\succ=2\pi k\rho_{0}.$ The effective field $B_{eff}$ is reduced
to
\begin{equation}
B_{eff}=B-\prec B\succ=B-2\pi k\varrho_{0} \tag{3.31}%
\end{equation}
Using the analogy of $\varkappa$ with the magnetic field $B$ $(eB\rightarrow
2\varkappa)$ we can substitute $eB_{eff}\rightarrow2\varkappa_{eff}$ \ and
find with $\rho_{0}=\frac{\varkappa}{\pi}\nu$ and $\nu=1/(2p+1\ $\ a
fractional $E^{c}$ charge $\varkappa_{eff}=\varkappa/(2p+1.$ This relation for
$\varkappa_{eff}$ agrees with the composite quark interpretation explained in
chapter 3.2 and the rules derived from the Yukawa interaction Lagrangian in
chapter 3.3.

\textit{\ }\textbf{\ 3.5 Emergent SU(3) symmetry in the tangent bundle}

The Chern-Simon formalism for the fractional QHE with an emergent $U(1)$ gauge
group has been generalized for the inclusion of spin or for a bilayer quantum
Hall system \cite{Q7,Q8,Q9,Q10}. In \cite{Q7,Q8} the fractional QHE was
described by two-component Abelian Chern-Simon fields, one for each spin or
each layer, and magnetic fluxes are attached to electrons in their own layer
and to the opposite layer. A detailed description of this approach for the
bilayer QHE is given e.g. in \cite{H1}. Note that this abelian two-component
Chern-Simon approach has a $U(1)\otimes U(1)$ gauge symmetry, and the $SU(2)$
spin symmetry is broken from the external construction. In \cite{Q9} \ a
non-Abelian Chern-Simon approach for $SU(2)$ spin singlets and in \cite{Q10}
for the $U(1)\otimes SU(2)$ symmetry was constructed in which a non-Abelian
Chern-Simon field \cite{Q11} is used as the source of an emergent additional
gauge field which is invariant with respect to the $SU(2)$ transformation of
the wavefunction. We apply some of the results of these papers for a treatment
of the vertical TB Laplacian and the description of the way, in which emergent
\ color $SU_{c}(3)$ gauge fields can arise together with fractional
hypercharges by an $U_{em}(1)$ field.

Quarks can occupy three different isospin states: two states are arranged as
an isospin dublet with $I_{3}=1/2$ and $I_{3}=-1/2)$ and the third is the
isospin singlet $I_{3}=0.$ In the Laplacian of the product group
\textbf{\ }$\mathbf{E}^{c}(2)$\textbf{\ }$\otimes SU(2)$ one can assume that
in the ground state the three isospin states are degenerate and equally
active. Therefore the system possesses an underlying $SU(3$) symmetry and the
vertical part of the TB Laplacian shows certain analog features of a quantum
Hall three-layer system.

In the case of a three layer system the eigenfunctions of the Laplacian are
described by three separate eigen functions $g_{I}(u)$ ($I=1,2,3)$ with $u$ as
the variables of the group. The eigenfunctions of the vertical Laplacian at a
fixed spacepoint $x$\textbf{\ }are represented as a $SU(3)$ spinor which will
be later identified as the quark spinor in the internal color space of QCD. We
denote the three components by red (r), green (g) and blue (b)
\begin{equation}
g(u)=%
\begin{bmatrix}
g_{r}(u)\\
g_{g}(u)\\
g_{b}(u)
\end{bmatrix}
\tag{3.32}%
\end{equation}
which possesses the underlying symmetry of the color SU(3) group
\begin{equation}
g_{I}%
\acute{}%
(u)=\exp(i\omega_{A}\lambda_{A})g_{I}(u) \tag{3.33}%
\end{equation}
with $\omega_{A}=$ $\omega_{A}(u)$. The SU(3) group has 8 generators
$\mathbf{\lambda}_{A}$ presented as $3\otimes3$ Lie algebra matrixes with the
commutation rule
\begin{equation}
\lbrack\mathbf{\lambda}_{A},\mathbf{\lambda}_{B}]=f_{AB}^{C}\mathbf{\lambda
}_{C} \tag{3.34}%
\end{equation}
where $f_{AB}^{C}$ are the structure constants of the SU(3) group. The
normalization is
\begin{equation}
Tr(\mathbf{\lambda}_{A}\mathbf{\lambda}_{B})=2\delta_{AB} \tag{3.35}%
\end{equation}

In the last section we described in the Chern- Simon approach for the
Laplacian of the $E^{c}(2)$ group the existence of an emergent $U(1)$ gauge
field connected with fractional hypercharges. As explained above due to the
three isospin components of quark states an additional $SU_{c}(3)$ symmetry
arises which leads to emergent color $SU(3)$ gauge fields.

For simplification we consider the case of a vacuum state with aligned spin
and isospin states. Similar to that in the Abelian Chern-Simon approach in
section 3.4, an Abelian and a non-Abelian Chern-Simon term is used to describe
the attachment of charge density $\varrho(\xi)$ and color $SU_{c}(3)$ spin
density to the quarks. The variables of number density $\varrho(\xi)$ and
color $SU_{c}(3)$ spin density in the ground state $n=0$ are defined as :%
\begin{align}
\varrho(\xi)  &  =\Phi_{vert}^{\dagger}(\mathbf{\xi})\Phi_{vert}(\mathbf{\xi
}),\tag{3.36}\\
S_{A}(\mathbf{\xi})  &  =\Phi_{vert}^{\dagger}(\mathbf{\xi})\lambda_{A}%
\Phi_{vert}(\mathbf{\xi})\nonumber
\end{align}
We denote the emergent $SU_{c}(3)$ Chern-Simon gauge vector potential as
$G_{a}^{A}(\mathbf{\xi})$, $(a=0,1,2,A=1-8$) and the emergent $U_{em}(1)$
gauge potential\textit{\ }as $A_{a}(\mathbf{\xi).}$Using the analog approach
as in chapter 3.4 the coupling of $SU(3)$ spinor quark fields to the emergent
gauge fields $A_{a}(\mathbf{\xi)}$ and $G_{a}^{A}(\mathbf{\xi})$ is described
by the generalized bundle Laplacian%

\begin{align}
\mathcal{L}  &  =\Phi_{vert}^{\dagger}[i\partial_{0}\Phi-(G_{0}^{A}%
(\mathbf{\xi})+A_{0}(\mathbf{\xi}))\Phi_{vert}]-\Lambda_{E}^{%
\acute{}%
}+L_{CS}\nonumber\\
&  \tag{3.37}%
\end{align}
where $\ $%
\begin{align}
\Lambda_{E}^{%
\acute{}%
}  &  =%
{\displaystyle\sum\limits_{A=1}^{8}}
\Phi_{vert}^{\dagger}(\mathbf{\xi)}[(-i\frac{\partial}{\partial\xi^{1}%
}+\varkappa\xi^{2}-\lambda_{A}G_{1}^{A}(\mathbf{\xi})-A_{1}(\mathbf{\xi}%
))^{2}\nonumber\\
&  +(-i\frac{\partial}{\partial\xi^{2}}-\varkappa\xi^{1}-\lambda_{A}G_{2}%
^{A}(\mathbf{\xi})-A_{2}(\mathbf{\xi}))^{2}]\Phi_{vert}(\mathbf{\xi})
\tag{3.38}%
\end{align}
is the transformed effective bundle Laplacian\textit{\ }of the $E^{c}(2)$
group, $\mathcal{L}_{CS}$ is the non-Abelian Chern-Simon action of the group
$U_{em}(1$)$\otimes SU_{c}(3)$ (compare \cite{Q9,Q10,Q11})
\begin{align}
\mathcal{L}_{CS}  &  =\frac{1}{4\pi k_{2}}\varepsilon^{abc}[G_{a}^{A}%
\partial_{b}G_{c}^{A}+\nonumber\\
&  +\frac{1}{3}f_{ABC}G_{a}^{A}G_{b}^{B}G_{c}^{C}]\tag{3.39}\\
&  +\frac{1}{4\pi k_{1}}\varepsilon^{abc}A_{a}(\mathbf{\xi)}\partial_{b}%
A_{c}(\mathbf{\xi)}\nonumber
\end{align}
with $\partial_{a}=\partial/\partial\xi^{a}$. $a,b,c=0,1,2$ , $k_{1},k_{2}$
are integer defining the topological structure of the model. The field
strength for the $U_{em}(1)$ group in the $E^{c}$ manifold is defined as
\begin{equation}
B(\mathbf{\xi)}=\epsilon^{ab}\partial_{a}A_{b}(\mathbf{\xi)} \tag{3.40}%
\end{equation}
and for the $SU(3$) group
\begin{equation}
G^{A}(\mathbf{\xi)}=\epsilon^{ab}(\partial_{a}G_{b}^{A}(\mathbf{\xi)-}%
f_{ABC}G_{a}^{B}G_{b}^{C}) \tag{3.41}%
\end{equation}
The equation of motion can be obtained by the variation of the Lagrangian $L$
over $A_{0}(\mathbf{\xi})$ given by%

\begin{equation}
B(\mathbf{\xi})=\varepsilon^{ab}\partial_{a}A_{b}=2\pi k_{1}\varrho
(\mathbf{\xi}) \tag{3.42}%
\end{equation}
Variation over $G_{0}^{A}(\mathbf{\xi})$ yields the constraint
\begin{align}
G^{A}(\mathbf{\xi})  &  =\epsilon^{ab}(\partial_{a}G_{b}^{A}-f^{ABC}G_{a}%
^{B}G_{b}^{C})\tag{3.43}\\
&  =2\pi k_{2}S^{A}(\mathbf{\xi})\nonumber
\end{align}

The invariant Chern-Simon SU(3) color-magnetic field $G^{A}(\mathbf{\xi})$ is
according to (3.43) directly proportional to the color $SU(3)$ spin density
$S^{A}(\mathbf{\xi})$. An important property of the (2+1)D Chern-Simon
approach is that the large-scale physics of an incompressible $2D$ system
(this means that there is an energy gap above the ground state) is determined
purely by the Chern-Simon action $L_{SC}$ in (3.39) \cite{Q6}. Other
interaction terms in (3.37) are short range and invisible in the
large-distance scale. This leads to an interesting conclusion. The field
equation derived from the pure Chern-Simon action $L_{SC}$ in (3.39) is given
by
\begin{equation}
G^{A}(\mathbf{\xi})=0 \tag{3.44}%
\end{equation}
Since the Chern-Simon color-magnetic field $G^{A}(\mathbf{\xi})$ in the
large-scale limit of the variables of the $E^{c}$ group vanishes we find from
(3.43)\ for the averaged color $SU(3)$ spin density
\begin{equation}
\prec S^{A}(\mathbf{\xi})\succ=0 \tag{3.45}%
\end{equation}
The vanishing of the average $\prec S^{A}(\mathbf{\xi})\succ$ over the
variables $\mathbf{\xi}$ implies that the ground state is a color singlet.
This can be interpreted as a signature of quark confinement in the TB-QFT and
explains why only colorless quark-antiquark pairs and colorless bound
three-quark systems forming mesons and baryons exist in nature. This
surprising and unexpected result in this paper follows from general universal
physical principles in the vertical TB\ Laplacian independent on the
microscopic dynamics of quarks in QCD. Note that similar universal properties
in a quantum Hall system, such as the Hall conductivity, are known where the
theoretical understanding of properties are encoded into the large-scale
Chern-Simon Lagrangian which do not involve a detailed understanding of the
microscopic quantum mechanics of such systems \cite{Q6}. However the
prediction of quark confinement must be confirmed by a complete microscopic
analysis of QCD in the TB which is beyond the present paper but will be
studied later.

In such a way the $SU(3)$ color symmetry is hidden in the vertical TB
Laplacian of the group $E^{c}(2)$ arising as an emergent symmetry. To describe
the full dynamics in the TB we have to include the spacetime-depending
horizontal part $\Phi_{hor}$ of the wavefunctions; therefore, the multiplets
of quark fields in the TB depend both on the spacetime variables $x$ as well
as on the variables $u$ of the associated bundle of the group $\mathbf{E}%
^{c}(2)$\textbf{\ }$\otimes SU(2)$:
\begin{equation}
q_{f}(x,u)=%
\begin{bmatrix}
q_{fr}(x,u)\\
q_{fg}(x,u)\\
q_{fb}(x,u)
\end{bmatrix}
\tag{3.46}%
\end{equation}
where the subscripts $r,g,b$ label the color states (red, green, blue). This
means the number density $\varrho(x,\xi)$ and color $SU_{c}(3)$ spin density
$S_{A}(x,\mathbf{\xi})$ in the ground state $n=0$ depend both on the spacetime
variable $x$ as well as on variables $u$ of the representation of the
$\mathbf{E}^{c}(2)$\textbf{\ }$\otimes SU(2)$ group:%

\begin{align}
\varrho(x,u)  &  =q^{\dagger}(x,u)q(x,u)\tag{3.47}\\
S_{A}(x,u)  &  =q^{\dagger}(x,u)\lambda_{A}q(x,u)\nonumber
\end{align}
The dependence of $\varrho(x,u)$ and $S_{A}(x,u)$ on the spacetime variables
$x$ requires us to include the horizontal part of the wavefunctions and the
treatment of gauge potentials and quark fields in dependence on the spacetime
variables. The dependence of the field strength tensor of the gluon fields on
the spacetime variables $x$ is defined as
\begin{equation}
G_{\mu\nu}^{a}=\frac{\partial}{\partial x^{\mu}}G_{\nu}^{a}-\frac{\partial
}{\partial x^{\nu}}G_{\nu}^{a}+g_{s}f_{abc}G_{\mu}^{b}G_{\nu}^{c} \tag{3.48}%
\end{equation}
with $g_{s}$ as the coupling coefficient for strong interaction. We can
combine the left-handed (L) and right-handed (R) quark fields into Dirac
spinors%
\begin{equation}
q_{f}=\binom{q_{Rf}}{q_{Lf}} \tag{3.49}%
\end{equation}
where $f=(u,d,c,s,t,b)$ is the flavour index $q_{Rf},$ $q_{Lf}$ \ refer to the
right or left handed particles. The Lagrangian density of the system is
$\mathcal{L}=\mathcal{L}_{q}+\mathcal{L}_{g}$ where $\mathcal{L}_{q}$ is the
Lagrangian density for the quark fields
\begin{align}
\mathcal{L}_{q}  &  =\sum\limits_{f}\overline{q}_{f}(x,u)[\gamma^{\mu}%
(i\frac{\partial}{\partial x^{\mu}}-m_{f}-\gamma^{0}\mu_{0}\nonumber\\
&  -g_{st}G_{\mu}^{A}(x)\mathbf{\lambda}^{A}]q_{f}(x,u). \tag{3.50}%
\end{align}
\emph{\ }The sum is over all flavours of quarks, $m_{f}$ is the quark mass and
$q_{f}(x,u)$ is a color triplet. The gluon Lagrangian density is given by%

\begin{equation}
\mathcal{L}_{g}=-\frac{1}{4}G_{\mu\nu}^{A}G^{A\mu\nu} \tag{3.51}%
\end{equation}
In the one-loop approximation the interaction Lagrangian density is determined
by%
\begin{equation}
\mathcal{L}_{int}=\frac{g_{st}^{2}}{2}%
{\displaystyle\int}
d^{4}yd\mu(u%
\acute{}%
)J^{A\mu}(x,u)D_{\mu\nu}^{AB}(x-y)J^{B\nu}(y,u%
\acute{}%
) \tag{3.52}%
\end{equation}
where $D_{\mu\nu}^{AB}(x-y)$ is the gluon propagator. The current is given by
\begin{equation}
J^{A\mu}=%
{\displaystyle\sum\limits_{f}}
\frac{1}{2}\overline{q_{f}}\gamma^{\mu}\mathbf{\lambda}^{A}q_{f} \tag{3.53}%
\end{equation}
From the Lagrange density one can obtain the Hamiltonian density via the
Legendre transformation. The Hamiltonian is obtained by integrating $h$ over
space $\mathbf{x}$ and the TB variables $u:$%

\begin{equation}
H(t)=%
{\displaystyle\int}
d^{3}xd\mu(u)[\pi(x,u)\frac{\partial q_{f}(x,u)}{\partial t}-\mathcal{L}(x,u)]
\tag{3.54}%
\end{equation}
where $d\mu(u)$ is the invariant measure of the group and
\begin{equation}
\pi(x,u)=\frac{\partial\mathcal{L}(x,u)}{\partial\frac{\partial q_{f}%
(x,u)}{\partial t}} \tag{3.55}%
\end{equation}
is the conjugant momentum of the field $q_{f}(x,u).$The above given equations
constitute the starting point for the construction of a generalized theory for
strong interaction based on the underlying geometry of the TB.

\textbf{4.} \textbf{The condensed vacuum structure in the TB, quark
condensation and the gap equation for quark-antiquark pairing }

\textbf{4.1 The completely filled vacuum state with family IQN }$n=0$

The understanding of the physical nature of the vacuum is a central problem in
modern quantum field theory rising some most profoundly mysterious unanswered
questions. Several well established factors hint that the vacuum -i.e. the
empty space- is in reality a complex structured medium. According the rules of
QED, the vacuum is not empty but actively populated by virtual
particle-antiparticle pairs that appear, annihilate and disappear during a
very short time. But even though in QED the vacuum is defined as the
zero-particle state in the Fock space, and the expectation values of quantized
matter fields and gauge fields are zero, due to the quantization these virtual
pairs have measurable effects in a shift of the spectrum of atomic systems and
the mass of elementary particles. In the SM of particle physics additional
phenomena arise which demonstrate that the vacuum may not be empty. The
classical Lagrangian of the Higgs field has a nonzero, constant value in its
lowest energy state, which means that the vacuum do not have zero field. This
resembles the Ginsburg-Landau model for superconductivity but up to now a
microscopic derivation of the Higgs mechanism has not been found which could
explain this phenomenon. The vacuum state of the Higgs field is a central
entity in the SM responsible for the spontaneous symmetry breaking of the
gauge symmetry and for the mass generation of gauge particles. In addition. in
QCD exist well established facts of the existence of a quark condensate
described by a non-vanishing vacuum expectation value of the composite quark
fields $\prec$ $\overline{q}q\succ$. This means that the vacuum is populated
by quark-antiquark pairs leading to measurable effects in QCD and connected
with a finite density of vacuum quark-antiquark pairs.

A further open question is whether the vacuum contains any energy. The
zero-point energy of all normal modes of quantized scalar or fermion fields
deliver\ a finite vacuum energy density. This has a profound effect for the
evolution of the universe, as the energy of the vacuum acts like a
cosmological constant in Einstein%
\'{}%
s equation of GTR and therefore it is \ a central entity for the evolution of
the universe. However, there is a huge discrepancy of 120 orders of magnitudes
between the theoretical prediction and the observed value obtained from
cosmological measurements. Moreover, the finite vacuum energy in QFT could be
connected with the dark energy problem in the expanding universe.

For the understanding of these problems significant changes in the
understanding of the vacuum in QFT seems to be required. Some outstanding
problems in elementary particle physics such as the Higgs condensate, chiral
quark condensation, fractional charges of quarks and the analogy of quark
confinement with the behavior of vortices in superconducting solid states seem
to find a counterpart in well understood phenomena in condensed matter
physics. However, these analogies arise in a heuristically phenomenological
way and there do not exists a consistent derivation of such relationships from
the fundamentals of the theory. Moreover significant differences exist between
particle physics and condensed matter physics, which seems to be an
insuperable obstacle to finding a basis for a consistent foundation of these
relationships. A central element in solid state theory is the existence of
well defined quantum energy bands, which differentiate between the ground
state and the excited states and arise due to the potential of periodic
arrangement of atoms or molecules in the solid state. The ground state in
semiconductors and isolators is a completely filled energy band denoted as the
Fermi sphere and is separated from the excited states by an energy gap. In the
SM of particle physics such difference between the ground state (vacuum) and
excited states do not exist; virtual particles in the vacuum are
differentiated from real particles only by the zero-particle number in Fock
space and can take energies continuously distributed up to a cut-off,
typically assumed to be the Plank energy.

In TB-QFT with the generalized gauge group $G=SU(2)\otimes$\textbf{\ }%
$E_{c}(2)$ significant conceptional chages arise in comparison with standard
QFT. As shown in chapter 3.1\ the Laplacian on the TB takes the form of the
multi-particle Hamiltonian of 2D electrons in an external magnetic field,
predicting two additional IQNs denoted here as the E$^{c}$-charge $\varkappa$
and the family quantum number $n$, which characterize different states
analogous to Landau levels for electrons in an external magnetic field. The
lowest number $n=0$ describes the vacuum, which means that the vacuum differs
from the excited states with different IQNs $n=1,2,3.$ The multi-electron
solid-state quantum Hall system shows a remarkable macroscopic quantum
phenomenon: the integer and the fractional quantum Hall effect (QHE).
Experimental observations indicate that such system undergoes a phase
transition into a peculiar ordered quantum state in which the Hall
conductivity is accurately quantized. In the integer QHE the ground state is a
completely filled level with a fixed and well-defined density, while in the
anomalous QHE the ground state is completely filled with composite electrons
which are bound objects of electrons paired with an even number of vortices.

The analog form of the vertical Laplacian with the Hamiltonian of a quantum
Hall system gives rise to the hypothesis that the vacuum with the IQN $n=0$ is
completely filled with seeleptons and composite seequarks, and all higher
levels $n=1,2,3$ are empty. In this picture, the background charge of vortices
cancel the charge with opposite sign of the composite quarks. When we add one
quark state to the system of the completely occupied ground state it is placed
into a higher energy level $n=1,2$,3 because of the Pauli exclusion principle
and leave an unoccupied state (hole) in the lowest energy state. This
resembles the understanding of an electron hole in a semiconductor crystal
lattice. In solid-state physics, an electron hole is simply the absence of an
electron from a full valence band. Similarly, the quark-hole introduced here
is a way to conceptualize the interaction of composite quarks (with $n=1,2,3$)
with the full vacuum state $n=0,$ which leads to a redefinition of
anti-particles as holes in the completely filled vacuum state.

The understanding of the vacuum as a completely filled band is a hypothesis
introduced in the present paper, but there exist similar concepts in physics.
Inspired by the exclusion principle Dirac proposed that the negative energy
states in his relativistic equation were in fact occupied states, and
excitations in this state can make holes interpreted as positrons. Even though
the Dirac interpretation was later revised, the status of a completely filled
ground state found in solid-state physics a deep foundation. A completely
filled solid-state band is inert, it remains a filled band at all times even
in the presence of space and time-dependend external fields. It cannot
contribute to electric and thermal currents, and only partially filled bands
contribute to conduction. Since a completely filled band carries no current,
an unoccupied state in a band evolve in time under the influence of external
fields precisely as it were occupied by real particles with opposite charge.
These fictitious particles are called holes in solid state theory, but in the
present meaning of the ground state as the vacuum, they have precisely the
same physical meaning as antiparticles in the standard QFT.

The new understanding of the vacuum as a completely filled band with the
family IQN $n=0$ differs significantly from the interpretation of the vacuum
in QFT as a Fock state with zero-particle number. Rather, it associates a
finite particle number density or a chemical potential with the vacuum state.\ 

\textbf{\ }A very simple model of a completely filled vacuum state could be
that of an ideal Fermi gas characterized by its density $\varrho=N/V$ where
$N$ is the total number of particles that occupy the lowest band and $V$ the
volume. The ideal Fermi gas is characterized by the Fermi energy $E_{F}$ or by
the Fermi momentum $k_{F}.$ The particle number density of the vacuum is
related with $k_{F}$ to
\begin{subequations}
\begin{equation}
\varrho_{vac}=g\frac{k_{F}^{3}}{3\pi^{2}} \tag{4.1}%
\end{equation}
where $g$ is the degeneracy factor given by the spin and internal quantum
number degree of freedom in the vacuum. Neglecting interactions, the Fermi
energy $E_{F}$ is approximately equal the chemical potential of the vacuum
$E_{F}\simeq\mu_{vac}.$

\textbf{4.2 Vacuum quark-antiquark condensation in QCD}

In the model of an ideal Fermi gas the interaction of particles is neglected.
In the presence of attractive interaction by gluon exchange, the vacuum is
instable with respect to the formation of a quark condensate due to the
pairing of quarks with antiquarks. This phenomenon shows some analogies with
superconductivity in solid states where pairs of electrons form a condensate
due to the weak attractive force mediated by phonons. Many phenological
observations leaves no doubt that in QCD a vacuum condensate of
quark-antiquark pairs is formed dynamically, leading to spontaneous breakdown
of the chiral symmetry. The quark condensate preserves Lorentz invariance of
the vacuum, but it does not preserve the global chiral symmetry. \emph{\ }

An early understanding of the formation of a quark condensate and chiral
symmetry breaking was achieved by the Nambu-Jona- Lasino model \cite{V12}
based on an interaction potential with delta-function behavior. According to
this model the formation of a quark condensate occurs similar as in the
microscopic theory of superconductivity in solid states developed by Bardeen,
Cooper and Schriefer \cite{B1}, where superconductivity is explained by the
formation of a condensate of electron-electron pairs (Cooper pairs). In
contrast the chiral condensate in QCD is formed by a condensate of
quark-anti-quark pairs.

The quark condensate is responsible for the spontaneous breaking of the chiral
symmetry in QCD. The order parameter for chiral symmetry breaking is given by
the vacuum expectation value $\prec0\mid\overline{q}q\mid0\succ$ . The value
of this parameter can be estimated by the Gell-Mann-Oakes-Renner relation
\cite{B2}
\end{subequations}
\begin{subequations}
\begin{align}
&  \prec0\mid\overline{q}q\mid0\succ=-\frac{f_{\pi}^{2}m_{\pi}^{2}}%
{2(m_{u}+m_{d})}\tag{4.2}\\
&  \simeq-(0.240)^{3}GeV^{3}\nonumber
\end{align}
Estimation of the quark condensate parameter can be directly done trough
analysis of correlation functions in QCD using an operator product expansion
where the value of the condensate is dictated by the value of the correlation
function \cite{B3}. In this approach QCD sum rules use the analytical
properties of current correlation functions to relate low-energy
(non-perturbative) hadronic quantities to calculable perturbative QCD
contributions at high energies.

The formation of a quark condensate by quark-antiquark pairing shows analogies
with a superconductor in solid state theory, but differ in one important
aspect. Electron-electron Cooper pairs responsible for superconductivity are
bosons and carry a charge 2e. In contrast quark-antiquark pairs are colorless
and are electrically neutral. Therefore they more resemble excitons in a
semiconductor which consists of electrons in the conduction band bound to
holes (realized as an unfilled electronic state in the valence band). The
possible condensation of excitons has been studied theoretically beginning in
the 1960th \cite{E1,E2,E3,E4}. Despite the formal similarity of an exciton
condensate and a superconductor, the physical properties of these two states
are very different. In an exciton condensate, the conductivity vanishes and
there is no Meisner effect and no long-range order like in superconductors.
Studies in semiconductor bilayer systems has provided experimental evidence
for the existence of exciton-condensation \cite{V4}.

Note that there exist an analog to solid-state superconductivity in QCD
characterized by the formation of bi-quark condensate in dense quark matter
(for reviews see e.g. \cite{V5,V6,V7}. At very high densities quark matter
could behave as a color superconducto, with possible phenomenological
applications for the description of neutron star interiors, neutron star
collisions or the physics of collapsing stars. While in these studies of color
superconductivity in QCD, extreme conditions of high-density quark matter are
assumed as e.g. in the interior of neutron stars in the context of the present
paper, dense quark matter is assumed to form the completely filled vacuum
state $n=0$ in the TB-QFT.

\textbf{4.3 The gap equation for quark-antiquark pairing}

Relativistic QFT is commonly formulated on the basis of Green functions using
the Lagrange density, which depends on the four spacetime coordinates $x^{\mu
}$ and includes retardation effects. To explore the dynamics of quark
condensation by quark-antiquark pairing we will apply in the mean-field
approximation a relativistic Hamiltonian formalism describing the two-body
interaction of quarks in the TB by gluon exchange.

The TB-vacuum is characterized as a completely filled state with a finite
particle density; therefore we have to consider a contribution to the
Hamiltonian arising from the finite particle number term $\mu N$ where
$\mu\simeq E_{Fermi}$ is the chemical potential and $E_{Fermi}$ the
Fermienergy of the vacuum which has to be determined later. The Hamiltonian
density (3.54) can be obtained from the Lagrangian density (3.50) via the
Legendre transformation$.$The Hamiltonian in the single-gluon exchange
approximation takes the form%

\end{subequations}
\begin{equation}
H=H_{0}+H_{I} \tag{4.3}%
\end{equation}
with%
\begin{align}
H_{0}  &  =%
{\displaystyle\int}
d^{3}xd\mu(u)\overline{q}_{f}^{a}(x,u)\tag{4.4}\\
&  \times\lbrack\gamma^{\mu}(\mathbf{\partial}_{\mu}-m-\gamma^{0}\mu_{0}%
]q_{f}^{a}(x,u)\nonumber
\end{align}
and
\begin{align}
H_{I}  &  =\frac{g^{2}}{2}%
{\displaystyle\int}
{\displaystyle\int}
d^{3}xd^{4}yd\mu(u)d\mu(u%
\acute{}%
)\tag{4.5}\\
&  \times J^{A\mu}(x,u)D_{\mu\nu}(x-y)J^{A\nu}(y,u%
\acute{}%
)\}.\nonumber
\end{align}
Hereafter, the summation convention concerning repeated color and flavor
indices is used. $A,B=1\sim8$ are the indices of the Gell-Man matrices. The
single-gluon exchange is attractive in the color-singlet channel for
quark-antiquark interaction. In relativistic treatment physical entities
depend on the four-momentum $k^{\mu}=(k_{0},\mathbf{k}).$ The interaction
Hamiltonian (4.5) depending on the gluon Green function $D_{\mu\nu}(x-y)$ is
not instantaneous$.$ In dense quark matter as discussed later in detail, the
screening of gluons play an important role and dominant contributions to the
interaction are at the edge of the Fermi-sphere. This allows to neglect
retardation effects.

In order to avoid problems, arising from the mixing of color, flavor and Dirac
indices in the interaction Hamiltonian, it is more convenient to introduce
Fiertz transformations for the sum of the Gell-Mann matrices:
\begin{align}%
{\displaystyle\sum\limits_{i=1}^{8}}
(\lambda^{A})_{ab}(\lambda^{A})_{cd}  &  =\frac{16}{9}(\delta)_{ad}%
(\delta)_{bc}\tag{4.6}\\
&  -\frac{1}{3}%
{\displaystyle\sum\limits_{i=1}^{8}}
(\lambda^{A})_{ad}(\lambda^{A})_{bc}\nonumber
\end{align}
where the first term describes color singlets and the second color octet
terms.\textbf{\ }Since the vacuum contains only color singlets, \ we consider
only the first term in (4.6). Therefore we can substitute in (4.5)%

\begin{align}
J^{A\mu}(x,u)J^{A\nu}(y,u%
\acute{}%
)\}  &  =g^{2}\frac{4}{9}[\overline{q}_{f}^{a}(x,u)\gamma^{\mu}q_{f}%
^{b}(x,u)\nonumber\\
\times &  \overline{q}_{g}^{b}(y,u%
\acute{}%
)\gamma^{\nu}q_{g}^{a}(y,u%
\acute{}%
)] \tag{4.7}%
\end{align}
The Fourier expansion of the quark field operator is determined by\emph{\ }%
\begin{align}
q_{f}^{a}(x,u)  &  =%
{\displaystyle\sum\limits_{\mathbf{p,}s}}
\frac{1}{\sqrt{2\epsilon_{p}V}}[e^{i\mathbf{px}}a_{fs}^{a}(\mathbf{p}%
)u_{s}(p)h_{f,s}(z)+\nonumber\\
&  +e^{-i\mathbf{px}}b_{fs}^{a\dagger}(\mathbf{p})v_{s}(\mathbf{p}%
)h_{f,s}^{\ast}(z)] \tag{4.8}%
\end{align}
where $f$ denotes the flavor degree of freedom with the IQN $M_{f}%
=$\{$n,m,j_{3},\varkappa\}.$ The hole (anti-quark) is an unfilled state in the
lowest TB-LL $n=0$ and the quark state is in an excited state with the IQN
n=1,2 3. $s=R,L$ is the helicity, $u_{s}(p)$\ with $p=(\epsilon(\mathbf{p}%
),\mathbf{p})$ is the plane-wave solution of the Dirac equation for a particle
and $v_{s}(p)$\ that of an antiparticle, and $h_{f,s}(z)$\ are the
eigenfunctions of the Laplacian of the $SU(2)\otimes E^{c}(2)$\ group. $V$ is
the volume of the system. The annihilation operators $a_{f,s}^{a}(\mathbf{p}%
)$\ and $b_{f,s}^{a}(\mathbf{p})$\ for particles and antiparticles (holes)
satisfying the anti-commutation relations destroy the unpaired vacuum
$\mid0\succ$, i.e. $a_{f,s}^{a}(\mathbf{p})\mid0\succ=b_{f,s}^{a}%
(\mathbf{p})\mid0\succ=0.$

In the absence of spin-orbit coupling we can assume that only spin-singlet
condensation takes place and spin-triplet pairing can be neglected. The
subsystem of particles of \ left-handed quarks (and right-handed anti-quarks)
can be treated independently from the subsystem of right-handed quarks (and
left-handed anti-quarks). Both are dynamically decupled and may be considered
as a non interacting mixture. This allow to simplify the notations and to drop
the spin-index.

Substituting (4.8) \ into (4.4)\ we obtain
\begin{align}
H_{0}  &  =%
{\displaystyle\sum\limits_{\mathbf{k}}}
(\epsilon_{q}(\mathbf{p})-\mu_{q})(a_{f}^{\dagger a}(\mathbf{p)}a_{f}%
^{a}(\mathbf{p)}+\nonumber\\
&  +(\epsilon_{h}(\mathbf{p})-\mu_{h})(b_{f}^{\dagger a}(\mathbf{p)}b_{f}%
^{a}(\mathbf{p)} \tag{4.9}%
\end{align}
where $\epsilon_{q}(\mathbf{p})$ is the kinetic energy of a quark and
$\epsilon_{h}(\mathbf{p})$ that of the holes, $\mu_{q}$ and $\mu_{h\text{ }}$
are the chemical potentials of quarks and holes, respectively. For the
interaction Hamiltonian (4.5) we obtain%
\begin{align}
H_{I}  &  =\frac{1}{2V}%
{\displaystyle\sum\limits_{\mathbf{p}_{1}\mathbf{p}_{2}\mathbf{p}%
_{3}\mathbf{p}_{4}}}
[V_{p_{1}p_{2}p_{3}p_{4}}^{qq}a_{f}^{a\dagger}(\mathbf{p}_{1})a_{g}^{b\dagger
}(\mathbf{p}_{2})\nonumber\\
&  a_{g}^{b}(\mathbf{p}_{3})a_{f}^{a}(\mathbf{p}_{4}\mathbf{)}\tag{4.10}\\
&  +V_{p_{1}p_{2}p_{3}p_{4}}^{hh}b_{f}^{a\dagger}(\mathbf{p}_{1}%
)b_{g}^{b\dagger}(\mathbf{p}_{2})b_{g}^{b}(\mathbf{p}_{3})b_{f}^{a}%
(\mathbf{p}_{4}\mathbf{)}\nonumber\\
&  -2V_{p_{1}p_{2}p_{3}p_{4}}^{qh}a_{f}^{a\dagger}(\mathbf{p}_{1}%
)b_{g}^{b\dagger}(\mathbf{p}_{2})b_{g}^{b}(\mathbf{p}_{3})a_{f}^{a}%
(\mathbf{p}_{4})]\nonumber
\end{align}
\ where $\mathbf{p}_{2}\mathbf{-p}_{3}=\mathbf{p}_{4}\mathbf{-p}%
_{1}=\mathbf{q}$. The interaction matrix elements are given by%

\begin{align}
V_{p_{1}p_{2}p_{3}p_{4}}^{qq}  &  =\frac{4}{9}g^{2}D_{\mu\nu}(q)K^{\mu}%
(p_{1},p_{4})K^{\nu}(p_{2},p_{3})J_{1}^{qq}\nonumber\\
V_{p_{1}p_{2}p_{3}p_{4}}^{hh}  &  =\frac{4}{9}g^{2}D_{\mu\nu}(q)Q^{\mu}%
(p_{1},p_{4})Q^{\nu}(p_{2},p_{3})J_{1}^{hh}\nonumber\\
V_{p_{1}p_{2}p_{3}p_{4}}^{qh}  &  =\frac{4}{9}g^{2}D_{\mu\nu}(q)[J_{1}%
^{qh}K^{\mu}(p_{1},p_{4})Q^{\nu}(p_{2},p_{3})\nonumber\\
&  -J_{2}^{qh}L^{\mu}(p_{1},p_{3})L^{\nu}(p_{2},p_{4})] \tag{4.11}%
\end{align}
with
\begin{align}
K^{\mu}(p_{1},p_{4})  &  =\overline{u(p_{1})}\gamma^{\mu}u(p_{4})\frac
{1}{2\sqrt{\epsilon_{\mathbf{p}_{1}}\epsilon_{\mathbf{p}_{4}}}},\nonumber\\
Q^{\mu}(p_{1},p_{4})  &  =\overline{v(p_{1})}\gamma^{\mu}v(p_{4})\frac
{1}{2\sqrt{\epsilon_{\mathbf{p}_{1}}\epsilon_{\mathbf{p}_{4}}}},\nonumber\\
L^{\mu}(p_{1},p_{3})  &  =\overline{u(p_{1})}\gamma^{\mu}v(p_{3})\frac
{1}{2\sqrt{\epsilon_{\mathbf{p}_{1}}\epsilon_{\mathbf{p}_{3}}}}. \tag{4.12}%
\end{align}
The initial 4-momenta are taken on the mass-shell. We also introduced the
integrals over the variables of the gauge group $G=SU(2)\otimes$%
\textbf{\ }$E_{c}(2)$ :%
\begin{equation}
J_{1}^{\varrho\sigma}=%
{\displaystyle\int}
d\mu(z)d\mu(z%
\acute{}%
)\mid h_{f}^{\varrho}(z)\mid^{2}\mid h_{g}^{\sigma}(z%
\acute{}%
)\mid^{2}, \tag{4.13}%
\end{equation}

\begin{equation}
J_{2}^{qh}=%
{\displaystyle\int}
d\mu(z)d\mu(z%
\acute{}%
)h_{f}^{q\ast}(z)h_{f}^{h\ast}(z)h_{g}^{q}(z%
\acute{}%
)h_{g}^{h}(z%
\acute{}%
), \tag{4.14}%
\end{equation}
with $\varrho,\sigma=q,h.$ If we excite a certain number of fermionic
particles from the fully occupied ground state across the Fermi surface, an
equal number of holes below the Fermi surface is obtained. Since the
eigenfunctions $h_{f}^{h}(z)$ for a hole with $n=0$ and $h_{f}^{q}(z)$ for an
excited valence quark with $n=1,2,3$ are orthogonal and normalized, we obtain
for the overlap integral in (4.13),(4.14) $J_{1}^{\varrho,\sigma}=1,J_{2}%
^{qh}=0.$ In the Feynman gauge the gluon propagator is given by
\begin{equation}
D^{\mu\nu}(q)=\frac{P_{T}^{\mu\nu}}{q^{2}-G_{T}(q)}+\frac{P_{L}^{\mu\nu}%
}{q^{2}-G_{L}(q)}, \tag{4.15}%
\end{equation}
where $P_{T,L}^{\mu\nu}$ are the transverse and the longitudinal projectors
\begin{align}
P_{ij}^{T}  &  =\delta_{ij}-\frac{q_{i}q_{j}}{\mid\mathbf{q}\mid},P_{00}%
^{T}=P_{0j}^{T}=0,\nonumber\\
P_{\mu\nu}^{L}  &  =-g_{\mu\nu}+\frac{q_{\mu}q_{\nu}}{q^{2}}-P_{\mu\nu}%
^{T}\nonumber\\
&  \cong\delta_{\mu0}\delta_{\nu0}. \tag{4.16}%
\end{align}
The entities $G_{T}(q)$ and $G_{L}(q)$ describe the screening of gluons in the
dense quark matter. In the dense hard loop approximation one can use the
approximations $G_{T}(p-q)=\frac{\mid\epsilon_{\mathbf{p}}\mathbf{-\epsilon
_{q}}\mid}{\mid\mathbf{p-q}\mid}N_{f}\frac{g^{2}\mu^{2}}{8\pi}\ $and
$G_{L}(p-q)=N_{f}\frac{g^{2}\mu^{2}}{2\pi^{2}}$ \cite{V8, V9}.

Similar as in the BCS theory or in the theory of exciton condensation, we
apply the mean-field approach for particle-particle interaction and
particle-antiparticle pairing. In this approximation mean-field decupling can
be used to generate terms containing the VEV of two of the four operators in
the interaction Hamiltonian (4.10). In such approximation we use
\begin{align}
&  \prec a_{f}^{a\dagger}(\mathbf{k})a_{g}^{a}(\mathbf{p})\succ=n_{f}%
^{q}(\mathbf{k})\delta_{\mathbf{kp}}\mathbf{\delta}_{fg}\tag{4.17}\\
&  \prec b_{f}^{a\dagger}(\mathbf{k})b_{g}^{a}(\mathbf{p})\succ=n_{f}%
^{h}(\mathbf{k})\delta_{\mathbf{kp}}\mathbf{\delta}_{fg}\nonumber
\end{align}
\ \ Taking into account only the pair attractive interaction with opposite
momenta $\mathbf{k,}$ the vacuum expectation for quark-hole pairing is
\textbf{\ }%
\begin{align}
&  \prec b_{g}^{b}(\mathbf{k)}a_{f}^{a}(\mathbf{k)\succ=}F_{gf}^{ba}%
(k)\tag{4.18}\\
&  \prec a_{f}^{b\dagger}(\mathbf{k})b_{g}^{b\dagger}(\mathbf{k)\succ=}%
F_{gf}^{ba\ast}(k)\nonumber
\end{align}
\ With this approximation the quark-quark interaction term in (4.10) is given
by
\begin{align}
H^{qq}  &  =\frac{1}{2V}%
{\displaystyle\sum\limits_{\mathbf{pk}}}
V_{pkpk}^{qq}[n_{f}^{q}(\mathbf{k})a_{g}^{b\dagger}(\mathbf{p})a_{g}%
^{b}(\mathbf{p)}\nonumber\\
&  -V_{pkkp}^{qq}n_{f}^{q}(\mathbf{k})a_{g}^{b\dagger}(\mathbf{p})a_{g}%
^{b}(\mathbf{p)}\nonumber\\
&  -V_{pkkp}^{qq}n_{f}^{q}(\mathbf{k})n_{g}^{q}(\mathbf{p})] \tag{4.19}%
\end{align}
A corresponding expression can be derived for the hole Hamiltonian $H^{hh}.$
The first term in (4.19)\ describes direct (Hartree) interaction and the
second the exchange interaction. Note that the Hartree terms of quarks and
holes are cancelled by the corresponding terms in the quark-hole interaction
Hamiltonian $H^{qh}$ which guarantees the charge neutrality. Therefore in
$H^{qh}$ we can include only the quark-hole pairing term%

\begin{align}
H^{qh}  &  =-\frac{1}{V}%
{\displaystyle\sum\limits_{\mathbf{pk}}}
V_{pkkp}^{qh}[F_{gf}^{ba}(k)a_{f}^{b\dagger}(\mathbf{p})b_{g}^{b\dagger
}(\mathbf{p)}\nonumber\\
&  +F_{gf}^{ba\ast}(k)b_{g}^{a}(\mathbf{p)}a_{f}^{b}(\mathbf{p})\tag{(4.20)}\\
&  -F_{fg}^{ab\ast}(k)F_{fg}^{ab}(p)]\nonumber
\end{align}
With local charge neutrality in the vacuum we have
\begin{align}
n^{q}(\mathbf{k})  &  =\prec a_{f}^{a\dagger}(\mathbf{k})a_{f}^{a}%
(\mathbf{k})\succ\tag{4.21}\\
&  =n^{h}(\mathbf{k})=\prec b_{f}^{a\dagger}(\mathbf{k})b_{f}^{a}%
(\mathbf{k})\succ\nonumber
\end{align}
Including the exchange energy into renormalized chemical potentials $\mu
_{ren}^{q}=\mu^{q}+\frac{1}{2}%
{\displaystyle\sum\limits_{\mathbf{k}}}
(V_{pkkp}^{qq}n^{q}(\mathbf{k}),\mu_{ren}^{h}=\mu^{h}+\frac{1}{2}%
{\displaystyle\sum\limits_{\mathbf{k}}}
(V_{pkkp}^{hh}n^{h}(\mathbf{k})$ and neglecting terms quadratic in
fluctuations, the Hamiltonian (4.10) takes the quadratic form%

\begin{align}
H  &  =%
{\displaystyle\sum\limits_{\mathbf{k}}}
[(\epsilon_{q}(k)-\mu_{ren}^{q})a_{f}^{a\dagger}(\mathbf{k})a_{f}%
^{a}(\mathbf{k})\tag{4.22}\\
&  +[\epsilon_{h}(k)-\mu_{ren}^{h})b_{f}^{a\dagger}(\mathbf{k})b_{f}%
^{a}(\mathbf{k})\nonumber\\
&  -%
{\displaystyle\sum\limits_{\mathbf{k,p}}}
V_{pkkp}^{qh}[F_{gf}^{ba}(\mathbf{k})a_{f}^{b\dagger}(\mathbf{p}%
)b_{g}^{a\dagger}(\mathbf{p)}\nonumber\\
&  +F_{gf}^{ba\ast}(\mathbf{k})b_{g}^{a}(\mathbf{p)}a_{f}^{b}(\mathbf{p}%
)+E_{vac}\nonumber
\end{align}
with the vacuum energy%
\begin{align}
E_{vac}  &  =-\frac{1}{V}%
{\displaystyle\sum\limits_{\mathbf{k},\mathbf{p}}}
V_{pkkp}^{qq}n^{q}(\mathbf{k})n^{q}(\mathbf{p})\nonumber\\
&  +V_{pkkp}^{hh}n^{h}(\mathbf{k})n^{h}(\mathbf{p})\tag{4.23}\\
&  +F_{fg}^{ab\ast}(\mathbf{k})F_{fg}^{ab}(\mathbf{p})]\nonumber
\end{align}
In order to diagonalize the Hamiltonian, we perform a Bogoljubov-Valentin
transformation and introduce new fermi-operators $A_{f}^{a}(\mathbf{k})$ and
$B_{f}^{a}(\mathbf{k})$ by the following linear combination for the operators
$a_{f}^{a}(\mathbf{k})$ and $b_{f}^{a}(\mathbf{k})$%
\begin{align}
a_{f}^{a}(\mathbf{k})  &  =u(\mathbf{k})A_{f}^{a}(\mathbf{k})+v(\mathbf{k}%
)B_{f}^{a\dagger}(\mathbf{k})\nonumber\\
b_{f}^{a}(\mathbf{k})  &  =-v(\mathbf{k})A_{f}^{a\dagger}(\mathbf{k}%
)+u(\mathbf{k})B_{f}^{a}(\mathbf{k}) \tag{4.24}%
\end{align}
The Bogoljubov-Valentin transformation offers a simple intuitive physical
insight for the physics of quark condensation and explicitly displays the
quark-antiquark pairing mechanism that is responsible for chiral symmetry breaking.

The Bogoljubov-Valentin transformation (4.24) preserves the fermionic
anti-commutation rule if \ the coefficients $u(\mathbf{k})$ and $v(\mathbf{k}%
)$ satisfy the condition
\begin{equation}
u(\mathbf{k})^{2}+v(\mathbf{k})^{2}=1 \tag{4.25}%
\end{equation}
Taking into account the pairing effect the ground state is now defined as
$A_{f}^{a}(\mathbf{k})\mid\Phi_{0}\succ=A_{f}^{a}(\mathbf{k})\mid\Phi_{0}%
\succ=0.$ The mean quark and hole occupation numbers are then given by \
\begin{equation}
n^{q}(\mathbf{p})=n^{h}(\mathbf{p})=v^{2}(\mathbf{p}) \tag{4.26}%
\end{equation}
satisfying the charge neutrality condition. By the request that terms
proportional to $A_{f}^{a}(\mathbf{k})B_{f}^{a}(\mathbf{k)}$ and
$B_{f}^{\dagger a}(\mathbf{k)}A_{f}^{\dagger a}(\mathbf{k})$ vanish, we obtain
the condition
\begin{equation}
2\xi(\mathbf{k})v(\mathbf{k})u(\mathbf{k})+\Delta(\mathbf{k)(}v(\mathbf{k}%
)^{2}-u(\mathbf{k})^{2})=0 \tag{4.27}%
\end{equation}
where the entity
\begin{equation}
\xi(\mathbf{k})=[(\epsilon(\mathbf{k})-\mu)-\frac{1}{\Omega}%
{\displaystyle\sum\limits_{\mathbf{k}}}
(V_{pkkp}^{qq}+V_{pkkp}^{hh})v(\mathbf{k})^{2} \tag{4.28}%
\end{equation}
is introduced with $\mu^{ren}=\mu_{q}^{ren}+\mu_{h}^{ren},\epsilon
(\mathbf{p})=\epsilon_{q}(\mathbf{p})+\epsilon_{h}(\mathbf{p}).$ The relation
(4.25) together with (4.27)\ are both satisfied by the solution
\begin{align}
u(\mathbf{k})^{2}  &  =\frac{1}{2}(1+\frac{\xi(\mathbf{k})}{E(\mathbf{k}%
)}),\tag{4.29}\\
v(\mathbf{k})^{2}  &  =\frac{1}{2}(1-\frac{\xi(\mathbf{k})}{E(\mathbf{k}%
)}),\nonumber
\end{align}
where $E(\mathbf{k})$ is the quasi-particle energy $E^{2}(\mathbf{k})=\xi
^{2}(\mathbf{k})+\mid\Delta(\mathbf{k)\mid}^{2}.$

Now we define the gap parameters $\Delta_{gf}^{ba}(\mathbf{p)}$ as follows
\begin{equation}
\Delta_{gf}^{ba}(\mathbf{p)}\mathbf{=}\mathbf{-}\frac{2}{V}%
{\displaystyle\sum\limits_{\mathbf{k,p}}}
V_{pkkp}^{eh}F_{gf}^{ba}(\mathbf{k}) \tag{4.30}%
\end{equation}
We consider the chiral limit with vanishing quark masses $m_{u}=m_{d}=m_{s}=0$
(and all heavy quarks neglected). In QCD, the chiral symmetry is spontaneously
broken down to the vectorial flavour subgroup $H=SO(3)_{L+R}U(1)_{L+R}$ of
isospin and hypercharge generated by the vector currents. This means that in
the limit of massless quarks the different flavours obtain the same vacuum
expectation value $\prec0\mid\overline{u}u\mid0\succ=\prec0\mid\overline
{d}d\mid0\succ=\prec0\mid\overline{s}s\mid0\succ$ \ and pairing of quarks and
anti-quarks with different flavors can be neglected. In agreement with this
consideration the gap $\Delta_{fg}^{ab}(\mathbf{p)}$ in Equ. (4.30) is
independent on the flavor and color index. Then equation (4.30) takes the
form
\begin{equation}
\Delta(\mathbf{p)}=-\frac{N_{g}}{V}%
{\displaystyle\sum\limits_{\mathbf{k}}}
V_{pkkp}^{eh}\frac{1}{E(\mathbf{k})}\Delta(\mathbf{k)} \tag{4.31}%
\end{equation}
with $N_{g}=N_{c}N_{f}$ ,where $N_{c}$ is the number of colors and $N_{f}$
\ the number of flavors.

After substitution of (4.24) with (4.25) and (4.27) into (4.22) the
Hamiltonian is%

\begin{align}
H  &  =E_{vac}+%
{\displaystyle\sum\limits_{\mathbf{k}}}
[E(\mathbf{k})[A_{f}^{a\dagger}(\mathbf{k})A_{f}^{a}(\mathbf{k})\nonumber\\
&  +B_{f}^{a\dagger}(\mathbf{k})B_{f}^{a}(\mathbf{k})]. \tag{4.32}%
\end{align}
The vacuum energy $(4.23)$ is given by%

\begin{align}
E_{vac}  &  =\frac{N_{g}}{V}%
{\displaystyle\sum\limits_{\mathbf{k,p}}}
(\epsilon_{\mathbf{k}}-\mu)\mid v_{\mathbf{k}}\mid^{2}\nonumber\\
&  -V_{\mathbf{kppk}}(v_{\mathbf{k}}u_{\mathbf{k}}v_{\mathbf{p}}u_{\mathbf{p}%
}+v_{\mathbf{k}}^{2}v_{\mathbf{p}}^{2})\tag{4.33}\\
&  =\frac{N_{g}}{V}%
{\displaystyle\sum\limits_{k,}}
[(\epsilon_{\mathbf{k}}-\mu_{ren})\ \frac{1}{2}(1-\frac{\xi_{\mathbf{k}}%
}{E_{k}})-\frac{1}{2}\frac{\Delta_{\mathbf{k}}^{2}}{E_{\mathbf{k}}}]\nonumber
\end{align}
The gap equation (4.31) takes the form
\begin{align}
\Delta(k)  &  =-\frac{4}{9}g^{2}N_{g}%
{\displaystyle\int}
\frac{1}{(2\pi)^{3}}d^{3}\mathbf{p}W^{\mu\nu}(\mathbf{p},\mathbf{k}%
,\mathbf{k},\mathbf{p})\nonumber\\
&  D_{\mu\nu}(p-q)\frac{1}{E(\mathbf{q})}\Delta(\mathbf{q)} \tag{4.34}%
\end{align}
with $W^{\mu\nu}(\mathbf{p},\mathbf{k},\mathbf{k},\mathbf{p})=K^{\mu
}(\mathbf{p},\mathbf{k})Q^{\nu}(\mathbf{k},\mathbf{p}).$

The integral in (4.34) can be solved using analog approximations as in
\cite{V9,V10,V11} in the study of color superconductivity by bi-quark
condensation. The gap function does not depend on the orientation of
$\mathbf{k}$\textbf{. }Due to the factor $E^{-1}(\mathbf{q})$ in (4.34) for
large chemical potential $\mu$ the main contribution in the integral arise at
the renormalized Fermi surface $\mid\mathbf{k}\mid\simeq\mu_{ren}%
,\mid\mathbf{p}\mid\simeq\mu_{ren}.$ With

$\mid\mathbf{q}\mid$=$\mid\mathbf{k-p}\mid\simeq\sqrt{2}\mu_{ren}(1-\cos
\theta)^{1/2}$ and $\epsilon(\mathbf{k})\simeq\epsilon(\mathbf{q})\simeq
\mu_{ren},$ $\mid\epsilon(\mathbf{k})-\epsilon(\mathbf{q})\mid\ll\mu_{ren}$ we find%

\begin{align}
\Delta(\mathbf{p)}  &  =-b_{qh}^{2}%
{\displaystyle\int\limits_{0}^{\pi}}
d\cos\theta%
{\displaystyle\int_{\mu_{ren}-\delta}^{\mu_{ren}+\delta}}
\ \mu_{ren}^{2}dk\tag{4.35}\\
&  [\frac{Q_{T}}{(\mu_{ren}^{2}(1-\cos\theta)-(\epsilon_{\mathbf{p}%
}\mathbf{-\epsilon_{k})}^{2}+G(k)}+\nonumber\\
&  +\frac{Q_{L}}{(\mu_{ren}^{2}(1-\cos\theta)-(\epsilon_{\mathbf{p}%
}\mathbf{-\epsilon_{k}\ )}^{2}+F}]\frac{\Delta(k\mathbf{)}}{E(k)}\nonumber
\end{align}
with $b_{qh}^{2}=\frac{1}{9}g^{2}N^{f}N_{C}\frac{1}{(\pi)^{2}}$ and $k$%
=$\mid\mathbf{k}\mid.$Here we restricted the integration over $\mathbf{k}%
$\textbf{\ }to a region near the Fermi surface with a cut-off factor $\delta.
$ The matrix elements $Q_{T,L}$ are defined as \ \textbf{\ }%
\begin{equation}
Q_{T,L}=\{\overline{u}(\mathbf{k})\gamma^{\mu}(u(\mathbf{p}))(\overline
{v}(\mathbf{p})\gamma^{\nu}\ v(\mathbf{k}))P_{\mu\nu}^{T;L}\frac{1}%
{4\epsilon_{\mathbf{p}}\epsilon_{\mathbf{k}}} \tag{4.36}%
\end{equation}
Let us choose $\mathbf{p}$ into the z-direction with $p=\epsilon_{\mathbf{p}%
}(1,0,0,1)$ and $k=\epsilon_{\mathbf{k}}(1,\sin\theta,0,\cos\theta).$ For the
matrix elements (4.12) one gets
\begin{align}
K_{RR}^{\mu}(p,k)  &  =K_{LL}^{\mu}(p,k)=Q_{LL}^{\mu}(p,k)=Q_{RR}^{\mu
}(p,k)=\nonumber\\
&  =2\sqrt{\epsilon_{\mathbf{p}}\epsilon_{\mathbf{k}}}(\cos\theta/2,\sin
\theta/2,i\sin\theta/2,\cos\theta/2)\nonumber\\
K_{RL}^{\mu}(p,k)  &  =Q_{RL}^{\mu}(p,k)=0 \tag{4.37}%
\end{align}
The matrix elements $Q_{T,L}$ in (4.36) take the form
\begin{equation}
Q_{T}=\frac{1}{2}(3-\cos\theta),Q_{L}=\frac{1}{2}(1+\cos\theta) \tag{4.38}%
\end{equation}
In (4.35) one can use the approximations $\mid\mathbf{\ (\epsilon_{\mathbf{p}%
}\mathbf{-\epsilon_{k})}\mid\ll}\mu_{ren}$ and $\mid\mathbf{p-k\mid\simeq
}\sqrt{2}\mu_{ren}(1-\cos\theta)^{1/2}.$ In leading order we can solve the
integral over $\theta$ by setting in the numerator $\cos\theta\simeq1$ . For
the gap equation (4.35) we find%

\begin{align}
\Delta(p\mathbf{)}  &  =\frac{2}{3}b_{qh}^{2}%
{\displaystyle\int_{-\delta}^{\delta}}
\frac{\mu^{2}dk}{\sqrt{\Delta^{2}(k)+\xi(k)^{2}}}[\frac{3}{2}\ln(1+\frac
{8\pi^{2}}{N_{f}g^{2}})\nonumber\\
&  +\ln(1+\frac{64\pi\mu}{N_{f}g^{2}\mid(\epsilon_{\mathbf{p}}%
\mathbf{-\epsilon_{k})\mid}})]\Delta(k\mathbf{)} \tag{4.39}%
\end{align}
with $p=\mid\mathbf{p}\mid,k=\mid\mathbf{k}\mid.$Analogous as in the studies
of color superconductivity \cite{V9,V10,V11}, the integral equation (4.39) can
be rewritten as a differential equation. Using the same approach, an
approximate solution of (4.39) is given by%
\begin{equation}
\Delta(k_{0}\mathbf{)}\mathbf{=}\Delta_{0}\sin[\frac{2}{3}b_{qh}\ln(\frac
{c\mu}{\epsilon(\mathbf{k})}) \tag{4.40}%
\end{equation}%
\begin{equation}
\Delta_{0}=2c\mu\exp(-\frac{3\pi^{2}}{\sqrt{2}g}) \tag{4.41}%
\end{equation}
with%
\[
2c=\frac{512\pi^{4}}{g^{5}}(\frac{2}{N_{f}})^{5/2}.
\]
Using these solutions one can approximately calculate the vacuum energy (4.33)
as
\begin{align}
E  &  =\frac{N_{g}\mu^{2}}{2\pi^{2}}%
{\displaystyle\int_{-\delta}^{\delta}}
d\xi\lbrack\frac{1}{2}\xi(1-\frac{\xi}{E_{k}})-\frac{1}{2}\frac{\Delta
_{\mathbf{0}}^{2}}{\sqrt{\xi^{2}+\Delta_{0}^{2}}}]\nonumber\\
&  =-\frac{N_{g}}{16(\pi)^{2}}\mu^{2}\Delta_{0}^{2} \tag{4.42}%
\end{align}
In the standard BCS theory of superconductivity there exist an internal
ultraviolet cut-off related with the Debye frequency $\omega_{D}$ because the
interaction potential of electrons due to phonon exchange is only attractive
in the interval $\mid\epsilon(p)-\mu\mid\preceq\omega_{D}.$ In QCD, such
cut-off frequency is absent. However, in the weak-coupling approximation the
results given in (4.40),(4.41) do not depend on the momentum cut-off $\delta$.

\emph{\ }For a check of the results we compare the gap equation for
quark-anti-quark condensation with the gap in bi-quark condensation in color
superconductivity in dense quark matter. Several fully microscopic
calculations of the gap for color superconductivity have been reported in the
literature (see e.g. \cite{V9,V10,V11}) by using Green functions in the
Nambu-Gorkov formalism. Here we applied the different method as described
above: the mean-field approach using the Bogoljubov transformation in the
Hamiltonian formalism. In the mean-field approximation of the Hamiltonian
formalism for bi-quark condensation we find the same expression (4.40) and
(4.41) for the gap as for quark-antiquark pairing but with the substitution of
the pre-factor $b_{qh}^{2}$ in (4.39) by $b_{qq}^{2}=g^{2}/18\pi^{2}.$This
agrees with the results in \cite{V9,V10,V11}.

Finally let us compare the gap equation for the quark condensate with the
results of the simplified Nambu-Jona-Lasino model with a point-like
interaction potential \cite{V12}. In this approach the gluon Green function in
(4.34) has to be substituted by $D(q)=g^{2}/M_{g}^{2}$ (where $M_{g}$ is an
effective gluon mass). The gap equation (4.34) with this point-like Green
function is given by
\begin{equation}
\Delta(p)=G_{s}\frac{1}{(2\pi)^{3}}%
{\displaystyle\int}
d^{3}k\frac{\Delta(\mathbf{k})}{\sqrt{\xi(\mathbf{k)}^{2}+\Delta
^{2}(\mathbf{k})}} \tag{4.43}%
\end{equation}
with $G_{s}=\frac{8}{9M_{g}^{2}}$ $g^{2}N_{c}N_{f}.$ The integration in (4.43)
can be performed if we regulate the ultraviolet divergence by a cutoff
$\widetilde{\mu}$ and substitute $\Delta(\mathbf{k})=\Delta_{0}\ $\ by a
constant. This solution agrees with the Nambu-Jona-Lasino model. For
$\Delta_{0}/\mu_{ren}\ll1$ one gets from (4.43)
\begin{equation}
\Delta_{0}=2\widetilde{\mu}\exp(-\frac{8\pi^{2}}{G_{s}}) \tag{4.44}%
\end{equation}
\ Note that the gap has an exponential proportional to $1/g^{2},$ while\ the
exponential behave in (4.41) is like $1/g.$ Similar to the theory of color
superconductivity in dense quark matter, the dominant contribution to the
formation of quark-antiquark pairs comes from the collinear scattering through
the long-range magnetic gluon exchange \cite{V9}.

\bigskip\textbf{4.4} \textbf{Numerical estimations}

Finally, it is interesting to obtain a few numerical estimates. Due to the
analogy with the fractional QHE, we introduced in the TB geometry a chemical
potential for the vacuum state as a completely filled fermion state. We
understand $\mu_{vac}$ as a new constant of nature. For the determination of
the parameter $\mu_{vac}$ we can use the relations of the gap parameter and
the known quark condensation parameter in dependence on the gap.

The quark field operator $\Psi_{f,s}^{a}(x,u)$ in (4.8) can be expressed by
the new operators $A_{f,s}^{a}(\mathbf{p})$ and $B_{f,s}^{a}(\mathbf{p})$
defined in (4.24)$:$
\begin{align}
\Psi_{f,s}^{a}(x,u)  &  =%
{\displaystyle\sum\limits_{p}}
\frac{1}{\sqrt{2\epsilon_{p}V}}[e^{i\mathbf{px}}A_{f,s}^{a}(\mathbf{p}%
)U_{s}(p)h_{f,s}(z)\nonumber\\
&  +e^{-i\mathbf{px}}B_{f,s}^{a\dagger}(\mathbf{p})V_{s}(\mathbf{p}%
)h_{f,s}^{\ast}(z)] \tag{4.45}%
\end{align}
where the new spinor functions $U_{s}(p)$ and $V_{s}(p)$ are given by
\begin{align}
U_{s}(p)  &  =u(\mathbf{p})u_{s}(p)-v(\mathbf{p})v_{s}(p)\tag{4.46}\\
V_{s}(p)  &  =u(\mathbf{p})v_{s}(p)+v(\mathbf{p})u_{s}(p)\nonumber
\end{align}
which retain the orthogonality and normalization conditions. The new paired
ground state$\mid\Phi\succ$ is annihilated by $A_{f,s}^{a}(\mathbf{p)}$ and
$B_{f,s}^{a}(\mathbf{p)}$ and $\mid\Phi\succ$ can be constructed from the
unpaired vacuum $\mid0\succ$ as \textbf{\ }
\begin{equation}
\mid\Phi\succ=%
{\displaystyle\prod\limits_{s,f,a}}
((u(\mathbf{p})-v(\mathbf{p})a_{f,s}^{a\dagger}(\mathbf{p})b_{f,s}^{a\dagger
}(\mathbf{p}))\mid0\succ\tag{4.47}%
\end{equation}
Using (4.45) with (4.46) and (4.47), one can determine the quark condensate
parameter $C_{q}$ as
\begin{align}
C_{q}  &  =\prec\Phi\mid\overline{\Psi}_{f}^{a}(x,u)\Psi_{f}^{a}(x,u)\mid
\Phi\succ\tag{4.48}\\
&  =-\frac{N_{g}}{V}%
{\displaystyle\sum\limits_{\mathbf{p}}}
u(\mathbf{p})v(\mathbf{p})=-N_{g}\frac{1}{V}%
{\displaystyle\sum\limits_{\mathbf{p}}}
\frac{1}{2E_{\mathbf{p}}}\Delta(\mathbf{p})\nonumber\\
&  =-\frac{N_{g}\Delta_{0}}{2(2\pi)^{3}}%
{\displaystyle\int}
d^{3}\mathbf{p}\frac{1}{\sqrt{\mathbf{p}^{2}+\Delta_{0}^{2}}}\simeq
-N_{g}\Delta_{0}\frac{\mu_{ren}^{2}}{8\pi^{2}}\nonumber
\end{align}
where the normalization of the factor $h_{f,s}(z)$ is used. Using the gap
$\Delta_{0}$ from (4.41), $C_{q}$ from (4.48) and (4.4) we find for the
renormalized chemical potential $\mu_{ren}$ the approximate relation
\begin{equation}
\mu_{ren}^{3}=-C_{q}\frac{4\pi^{2}}{N_{g}c}\exp(\frac{3\pi^{2}}{\sqrt{2}g})
\tag{4.49}%
\end{equation}
We need the behavior of the fundamental coupling constant $g(Q^{2})$ in the
low momentum transfer domain $Q\prec1GeV.$ Various theoretical models in the
non-perturbative strongly coupled regime has been developed. Here we use the
effective coupling from light-front holography \cite{V12}, leading to a
non-perturbative coupling $\alpha_{s}(Q^{2})=\pi\exp(-Q^{2}/4\varkappa^{2})$
with $\varkappa=0,54GeV$ . For $Q=0.3GeV$ we find $g=\sqrt{4\pi\alpha_{s}%
}\simeq6.0.$ With this value we find a vacuum chemical potential $\mu
\simeq0.8GeV$ and a gap $\Delta_{0}=2c\mu_{ren}\exp(-\frac{3\pi^{2}}{\sqrt
{2}g})\simeq0.2GeV$. For the vacuum energy density we obtain $E_{vac}%
\simeq-10^{-3}GeV^{4}.$ Galactic observations deliver an upper bound on the
cosmological constant which is usually interpreted as a bound on the vacuum
energy in the order of $10^{-47}GeV^{4}.$ In contrast the zero-point energy in
QFT is roughly $10^{76}GeV^{4}$ if we insert the Plank energy as cut-of. This
huge difference of more than 120 orders constitutes the so-called cosmological
constant problem.

\emph{\ }With these rough estimations we want to emphasis the underlying
physics, but a more exact numerical will be undertaken in future.

\ 

\bigskip\textbf{5. Summary }

This paper is a follow-up of \cite{I1} with a study of the hypothesis that the
tangent bundle (TB) with the structure group $SO(3,1)\rtimes T(3,1)$ is the
underlying geometric structure for a unified theory of the fundamental
interactions, explaining their common origin and enabeling a deeper
understanding of the relationship between them. Based on this assumption in
\cite{I1} a generalized theory of electroweak interaction (including
hypothetical dark matter particles) with the little groups G = $SU(2)\otimes
E^{c}(2)$ of the $SO(3,1)$ group as gauge group was presented. The present
paper describes a possible way that strong interaction can emerge in the
tangent bundle geometry. The group $SU(3)$ cannot be described as a
geometrical symmetry in the TB but this symmetry is hidden in the fundamentals
of the tangent bundle geometry arising as an emergent internal symmetry
similar as Chern-Simon gauge symmetries in quantum Hall systems. This
assumption is based on the fact that the vertical Laplacian of the TB has the
same form as the multi-particle Hamiltonian of a quantum Hall system. The
eigensolutions of the vertical Laplacian exhibit two additional internal
quantum numbers (IQN) which explain the existence of lepton and quark
families: the $E^{c}$-charge $\varkappa$ and the family quantum number $n$.
The family quantum number $n$ characterizes different states analogous to
Landau levels of electrons in an external magnetic field. The lowest quantum
number describes a completely occupied vacuum state (filled with seeleptons
and seequarks). This means the vacuum state (with n=0) differs from excited
states (with $n=1,2,3$) describing valence quarks by a different IQN. \ The
analogy with a quantum Hall system allows us to use the Laughlin wave function
for the description of quarks with fractional hypercharges which can be
interpreted as composite quarks formed from bare quarks and two attached
hypercharge vortices. Taking into account the three iso-spin components
$I_{3}=-1/2,0,1/2$ the color SU(3) symmetry arises as an emergent gauge
symmetry described by (2+1)D Chern-Simon gauge fields. The field equation of
the vertical Laplacian including the emergent Chern-Simon fields implies that
in the large-scale limit of the variables of the TB Laplacian the ground state
is a color singlet demonstrating a signature of quark confinement. This result
follows from general universal principle in the TB vertical Laplacian
independent on the microscopic dynamics of quarks in QCD. In addition in the
TB geometry, a new understanding for the vacuum as the ground state is
introduced that is occupied by a condensate of quark-antiquark pairs with
finite density (or a chemical potential). The gap for quark-antiquark pairing
is calculated in the mean-field approximation, which allows a numerical
calculation of the characteristic parameters of the vacuum such as its
chemical potential, the chiral condensation parameter and the vacuum energy.

Recently at CERN new exotic particles were observed formed as tetraquarks
containing two quarks and two antiquarks and pentaquark containing four quarks
and one antiquark (for a review see e.g. \cite{V14}). The analogy with the
anomalous QHE could hint to the possible existence of other types of exotic
particles formed from exotic quark states with hypercharges of e/5 for up and
down quarks and the exotic up quark had an electric charge of (7/10)e. The
exotic down quark had a charge of (-3/10)e. Pairs of exotic quark and exotic
antiquarks with the same flavour can form neutral flavorless exotic mesons.
The real existence of e/5 charged quasiparticles has been proven by shot-noise
measurements in a quantum Hall system \cite{V15}.

Since the tangent bundle is also the geometric fundament for a gauge theory of
gravity based on translational transformations $T(3,1)$ of tangent fibers
\cite{t1,t2,t3,t4,t5} one can identify the TB as the underlying geometric
structure for a new type of unified geometrized field theory. However the
hardest unsolved problem in the unification of fundamental interactions is the
quantization of gravity. While there has exciting developments in this field,
to date the ongoing intense research efforts did not resulted in a general
accepted consistent theory. In the best case, the underlying unified geometric
structure of the TB with conceptual changes in comparison with GTR, as
described in this paper, could provide a new perspective to this problem to
allow a deeper understanding of the nature of quantum gravity.


\bigskip\textbf{References}

\begin{thebibliography}{99}                                                                                               %


\bibitem {I1}J. Herrmann, Eur. Phys. J. C 79, 779 (2019), ArXiv:1802.03228v3

\bibitem {G1}H. Georgi and S. Glashow, Phys. Rev. Lett. 32, 438 (1974),

\bibitem {G2}H. Fritsch and P. Minkowski, Ann. Phys. (N.Y.) 93, 193 (1975)

\bibitem {G3}F. Wilszek and A. Zee, Phys. Rev. Lett.42, 421 (1979)

\bibitem {G4}C. D. Froggatt and H.B.Nielsen, Nucl. Phys. B147, 277 (1979)

\bibitem {r1}E. Lubkin, Ann.Phys. (N.Y.) 23, 233 (1963),

\bibitem {r2}A. Trautman, Rep. Math. Phys. 1, 29 (1970),

\bibitem {r3}W. Drechsler W and M. E. Meyer; "Fibre Bundle Techniques in Gauge
Theories", Lecture Notes in Physics, Springer 1977,

\bibitem {r4}M.F. Atiyah "The geometry of Yang-Mills fields", Scuola Normale
Superiore, Pisa, Italy (1979),

\bibitem {r5}M. Daniel and C. M. Viallet, Rev. Mod. Phys. 52, 175 (1980),

\bibitem {r6}M. Nakahara, "Geometry, Topology and Physics", Institute of
Physics , Bristo 1990,

\bibitem {r7}Bo-Yuan Hou, "Differential geometry for physicists", Singapore 1997,

\bibitem {t1}K. Hayashi and T. Nakno, Prog Theor. Phys. 38, 491 (1967),

\bibitem {t2}Y. M. Cho, Phys. Rev. D14, 2521 (1976),

\bibitem {g5}T. Dass, Pramana \textbf{23}, 433 (1984),

\bibitem {t3}K. Hayashi and T. Shirafuji, Phys. Rev. D19, 3524 (1979),

\bibitem {t4}J. W. Maluf, Annalen Phys. 525, 339 (2013),arXiv:1303.3897 [gr-qc],

\bibitem {t5}R. Aldrovandi, j. G. Pereira, "Teleparallel gravity; an
introduction" (Springer, Dordrecht, 2012),

\bibitem {P1}T. W. Kibble, J. Math. Phys. 2, 212, (1961),

\bibitem {p2}D. W. Sciama, in "Recent Developements in General Relativity"
(Pergamon, New York, 19962) p.415

\bibitem {P2}D. Ivanenko and G. Sardanashvily, Phys. Repts. 94, 3 (1983),

\bibitem {P3}F.W. Hehl, J.D. McCrea, E.W. Mielke and Y. Ne%
\'{}%
eman, Phys. Rep. 258, 1 (1995),

\bibitem {P4}M. Blagojevic and F. W. Hehl (eds) "Gauge theories of
Gravitation. A reader with Commentaries", Imperial College Press. London 2013,

\bibitem {D1}W. Kopczynski, J. Phys.: Math. Gen. 15, (1982) 493

\bibitem {D2}M. Fontanini, E. Huguet and M. Le Delliou, Phys. Rev. D99, 064006 (2019),

\bibitem {D3}J. G. Pereira and Y. N. Obukhov, Universe 2019, 5, 139[8],

\bibitem {TG1}S. Kobayashi an K. Nomizu, \textquotedblright Foundation of
Differential Geometry\textquotedblright, Interscience Publishers (1963),

\bibitem {TG2}C. J. Isham, \textquotedblright Modern Differential Geometry for
Physicists\textquotedblright, World Scientific (1999),

\bibitem {r8}E. P. Wigner, Ann. Math. 40, 149 (1939),

\bibitem {H1}Z. F. Ezawa, \textquotedblright Quantum Hall
Effects\textquotedblright, World Scientific, 2013,

\bibitem {H2}S.C. Zhang, H.Hansson and S.Kivelson, Phys. Rev. Lett.62, 82, (1989),

\bibitem {H3}A. Lopez and E. Fradkin, Phys. Rev. B 44, 5246 (1991),

\bibitem {H4}O.Heinonen (ed.), \textquotedblright Composite
Fermions\textquotedblright, World Scienific, 1998,

\bibitem {H5}J. K. Jain, \textquotedblright Composite
Fermions\textquotedblright, Cambridge University Press, 2009

\bibitem {H6}R. B. Laughlin, Rev. Mod. Phys. 71, 863 (1999) - Nobel lecture,

\bibitem {E1}I. V. Keldysh and Y. V. Kopaev, Fiz. Twerd. Tela 6, 2791 (1964),

\bibitem {E2}D. Jerome, T. M. Rice and W. Kohn, Phys. Rev., 158, 452 (1967),

\bibitem {E3}C. Comte and P. Nozieres, Journal de Physique, 1982, 43 /7) 0.1069,

\bibitem {E4}S. A. Moskalenko and D. W. Snoke, "Bose-Einstein Condensation of
Excitons and Biexcitons ", Cambridge university press , Cambridge (2000),

\bibitem {r9}V. Bargman, Ann. Math. 59, 1 (1954,

\bibitem {r10}H. Hoogland, J. Phys. A Math. Gen. 11, 1557 (1978),

\bibitem {r11}J. F. Carina, M. A. del Olmo and M Santander, J. Phys. A:
Math.Gen. 17, 3091 (1984),

\bibitem {r12}S. Weinberg, "The quantum theory of fields" vol 1, Cambridge
University Press, Cambridge,

\bibitem {b1}I.M. Shapiro, R.A. Minlos and Z.Ya Shapiro, "Representations of
the rotation and Lorentz groups and their applications", Pergamon Press (1963),

\bibitem {QH1}K. v. Klitzing, G. Dorda and M. Pepper, Phys. Rev. Lett. 45, 494 (1980),

\bibitem {QH2}C.Tsui,H.L.Storme$\operatorname{rand}$A.C.Gossard, Phys. Rev.
Lett. 48, 1559 (1982),

\bibitem {QH3}R. B.Laughlin, Phys. Rev. Lett.50,1395 (1983),

\bibitem {QH4}B. Halperin, Helv. phys. acta 56, 75 (1983),

\bibitem {Q5}J. K. Jain, Phys. Rev. Lett. \textbf{63}, 199 (1989),

\bibitem {Q6}J. Fr\"{o}hlich, A. Zee, Nucl. Phys. B364, 517 (1991),

\bibitem {Q7}X. G. Wen and A. Zee, Phys. Rev. B 47, 2265 (1993),

\bibitem {Q8}A. Lopez and E. Fradkin, Phys. Rev. B51,4347 (1995),

\bibitem {Q9}A. Balatsky and E. Fradkin, Phys. Rev. B 43, 10622 (1991)

\bibitem {Q10}J. Fr\"{o}hlich, T. Kerler and P.A. Marchetti, Nucl. Phys. 374,
511 (1992),

\bibitem {Q11}E. Witten, Commun. Math. Phys. 121, 351 (1989),

\bibitem {V4}J. P. Eisenstein and A. N. MacDonald, Nature 432, 691 (2004),

\bibitem {V5}D. Bailin and A Love, Phys.Rep.103, 325 (1984),

\bibitem {V6}K.Rajagopal, F. Wilczek, "The condensed matter Physics of QCD" in
handbook of QCD, volume 3, 200,World Scientific, Singapore (2001),

\bibitem {V7}M. G. Alford, A. Schmitt.K. Rajagopal and T.Sch\"{a}fer,
Rev.Mod.Physics 80, 1455 (2008),

\bibitem {V8}Thermal Field Theory, Cambridge University Press, (Cambridge, 1996),

\bibitem {V9}D. T. Son, Phys. Rev. D59, 094019 (1999),

\bibitem {V10}T. Sch\"{a}fer and F. Wilczek, Phys. Rev. D60, 114033 (1999),

\bibitem {V11}R. D. Pisarski and D.H. Rischke, Phys. Rev. D61, 051501(R)(2000),

\bibitem {V12}Y.Nambu and G. Jona-Lasinio, Phys. Rev. 122, 345 (1961),

\bibitem {B1}J. Bardeen, L. N Cooper, and J. R. Schrieffer, Phys. Rev. 108,
1175 (1957),

\bibitem {B2}M.Gell-Mann, R. J. Oakes, and B. Renner, Phys. Rev. 175, 2195 ((1968)

\bibitem {B3}M. A. Shifman, A. I. Vainshtein and V. I. Zakharov, Nucl. Phys. B
147, 385, 448 (1979),

\bibitem {V13}S.J.Brodsky, A.Deur, and G.F.deTramond, arXiv:1002.3948v3[hep-ph],

\bibitem {V14}A. Ali, J.S. Lange and S. Stone arXiv:1706.00610v2,

\bibitem {V15}M. Reznikov et al, Nature 399, 238 (1999).
\end{thebibliography}
\end{document}